\documentclass[10pt,journal]{IEEEtran}

\usepackage{fancyhdr}

\fancypagestyle{firstpage}{
    \fancyhf{}
    \fancyhead[L]{%
        \footnotesize
        This work has been submitted to the IEEE for possible publication.
        Copyright may be transferred without notice, after which this version
        may no longer be accessible.
    }
    
}

\usepackage{amsmath} 
\usepackage{amssymb} 
\usepackage{indentfirst}
\usepackage{graphicx}
\IEEEoverridecommandlockouts
\usepackage{cite}
\usepackage{amsmath,amssymb,amsfonts}
\usepackage{algorithmicx}
\usepackage{algorithm}
\usepackage{algpseudocode}  
\usepackage{amsmath}  
\usepackage{graphicx}
\usepackage{textcomp}
\usepackage{xcolor}
\usepackage{float}
\usepackage{graphicx}
\usepackage[caption=false,font=footnotesize]{subfig}
\usepackage{verbatim}
\def\BibTeX{{\rm B\kern-.05em{\sc i\kern-.025em b}\kern-.08em
    T\kern-.1667em\lower.7ex\hbox{E}\kern-.125emX}}

\usepackage[caption=false]{subfig}
\usepackage{xcolor}
\usepackage{xpatch}
\makeatletter
\def\changeBibColor#1{%
  \in@{#1}{}
  \ifin@\color{blue}\else\normalcolor\fi
}

\xpatchcmd\@bibitem
  {\item}
  {\changeBibColor{#1}\item}
  {}{\fail}
 
\xpatchcmd\@lbibitem
  {\item}
  {\changeBibColor{#2}\item}
  {}{\fail}
\makeatother

\begin{document}

\title{Polarization-Aware Rotatable Antennas for RIS-Empowered Symbiotic Radios}

\author{ 
Chao Zhang, \emph{Graduate Student Member, IEEE}, Ruizhe Long, \emph{Member, IEEE}, \\Boon‑Hee Soong, \emph{Life Senior Member, IEEE}, and Ying-Chang Liang, \emph{Fellow, IEEE}
\\
\thanks{Part of this work has been accepted for presentation at the 2026 IEEE Global Communications Conference (GLOBECOM 2026)~\cite{Zhang2026conference}.}
\thanks{C.~Zhang and R.~Long are with the National Key Laboratory of Wireless Communications, University of Electronic Science and Technology of China, Chengdu 611731, China (e-mail: {zhang\_chao@std.uestc.edu.cn, ruizhelong@gmail.com}).}
\thanks{B. H. Soong is with the School of Electrical and Electronic Engineering, Nanyang Technological University, Singapore 639798 (e-mail: {ebhsoong@ntu.edu.sg}).}
\thanks{Y.-C. Liang is with the Institute for Fundamental and Frontier Sciences, University of Electronic Science and Technology of China, Chengdu 611731, China (e-mail: {liangyc@ieee.org}). }
}
\maketitle
\thispagestyle{firstpage}
\begin{abstract}
This paper investigates a dual-polarized reconfigurable intelligent surface (DP-RIS)-empowered symbiotic radio (SR) system with rotatable antennas (RAs). By reconfiguring antenna orientations, RAs can steer their radiation patterns toward desired directions, thereby effectively mitigating the double-fading effect in RIS-assisted cascaded links. However, antenna rotation not only changes the radiation direction but also alters the local polarization bases, which may result in polarization mismatch and degrade the achievable gain.
This motivates a joint spatial-polarization design that simultaneously exploits directional radiation gain and polarization matching. Specifically, we formulate a transmit power minimization problem that jointly optimizes digital beamforming, RA rotations, transceiver polarization states, and DP-RIS phase shifts, subject to the primary and secondary rate requirements as well as interference temperature constraints for non-SR users. To solve this non-convex problem, we develop an alternating optimization algorithm that integrates semidefinite programming, difference-of-convex programming, and Riemannian conjugate gradient methods. 
Moreover, to reduce hardware and computational complexity for practical deployment, we propose two low-complexity RA designs, namely a subarray-wise shared rotation design and a discrete rotation codebook design. Simulation results show that the proposed polarization-aware RA design significantly reduces the required transmit power compared with fixed-orientation and polarization-unaware benchmark schemes. Moreover, both the proposed low-complexity RA designs achieve comparable performance with reduced rotation complexity.
\end{abstract}

\section{Introduction}
Driven by the proliferation of massive connectivity and low-power Internet-of-Things (IoT) devices, achieving ultra-high spectral and energy efficiency has become critical for sixth-generation (6G) networks~\cite{asghar2022evolution,wang2023road}. Symbiotic radio (SR), which has evolved from cognitive radio and ambient backscatter communication, has emerged as a promising technology for improving both spectral and energy efficiency~\cite{long1,Wang2026symbiotic}. Specifically, SR enables backscatter devices (BDs) to transmit their information by modulating the primary signal without requiring dedicated radio-frequency (RF) chains or additional spectrum resources, while the reflected signal can also provide an additional propagation path for the primary transmission. Despite these appealing advantages, SR still faces two major challenges: severe double fading in the backscatter link and interference leakage to coexisting non-SR users.

To cope with the above challenges, reconfigurable intelligent surfaces (RISs) have been integrated into SR systems to serve as BDs and reshape the propagation environment~\cite{11012663,zhang2026multi}. Through joint active and passive beamforming, these RIS-empowered systems can effectively enhance secondary transmissions while suppressing interference to non-SR users. However, existing RIS-empowered SR designs mainly rely on beamforming to control the spatial energy distribution, while the radiation directions of base station (BS) antennas remain fixed after deployment. As a result, the directional gain of BS antennas cannot be flexibly adapted to the spatial distribution of these nodes. This limitation restricts the degrees of freedom in the antenna domain, thereby reducing the capability of existing designs to alleviate the double fading in the cascaded link and suppress interference leakage.

Recently, reconfigurable antenna technologies have attracted increasing attention for unlocking new degrees of freedom in the antenna domain beyond conventional fixed arrays~\cite{11203239}. Among them, movable antennas (MAs), fluid antenna systems (FASs), and six-dimensional movable antennas (6DMAs) improve channel conditions by reconfiguring antenna positions and orientations~\cite{MA1,MA2,6DMA,shao_20256dma}. Different from these position-reconfigurable architectures, rotatable antennas (RAs) keep antenna locations fixed and adjust only their orientations, thereby enabling orientation reconfigurability without altering the antenna array structure~\cite{zheng2026t}. Existing RA studies mainly exploit this flexibility to optimize spatial radiation characteristics, including directional gain enhancement, beam steering, and interference suppression. By jointly optimizing antenna orientations and beamforming, RAs have shown their effectiveness in various wireless scenarios, including multi-user communications~\cite{11427014}, secure transmission~\cite{11098736}, covert communication~\cite{11426957}, and integrated sensing and communication (ISAC)~\cite{RA}. 



However, antenna rotation affects not only the spatial radiation direction but also the local polarization bases~\cite{Joint1}. The resulting polarization mismatch may partially offset the directional gain achieved through orientation adjustment~\cite{zhang2026polarization}. Although only a limited number of studies have investigated such rotation-induced polarization effects, several initial efforts have demonstrated the importance of incorporating polarization into orientation-reconfigurable antenna design. In~\cite{PRA}, polarization-reconfigurable antennas have been investigated in a 6DMA-assisted ISAC system, where localization and joint optimization have been employed to demonstrate the performance advantages of 6DMAs. In~\cite{zhang2026polarization}, the impact of polarization mismatch on the directional gain of RAs has been examined, and the performance gains have been decomposed into three components: antenna directivity gain, polarization-orientation alignment gain, and polarization-state matching gain.
These studies have shown that polarization should be considered together with antenna rotation. In RIS-empowered SR systems, however, polarization design is not limited to the rotation-induced effects of RAs, since the effective gains of the direct and RIS-assisted links also depend on the transceiver polarization states and the polarization response of the RIS.
At the transceiver side, dual-polarized massive MIMO has been investigated to exploit the polarization domain for improving spectral efficiency~\cite{9894274}. To reduce the associated hardware overhead, polarization-reconfigurable antennas have been proposed to adjust polarization states with lower implementation complexity~\cite{PRA3}. More recently, polarforming has been developed to flexibly control the amplitude and phase of transmit polarization states using power splitters and phase shifters~\cite{PRA,PRA4}. On the RIS side, dual-polarized RISs (DP-RISs) have been investigated to support polarization-dependent reflection through two orthogonal polarization components. For example, DP-RIS architectures have been used to directly modulate and transmit multichannel signals without relying on conventional RF chains~\cite{9475466}. Moreover, DP-RIS configurations have been designed for broad-beam reflection and holographic MIMO systems, where the two polarization dimensions are exploited to improve coverage and multiplexing performance~\cite{10785256,Zeng2024dual}. Nevertheless, existing RA and polarization-aware studies have largely been developed in separate settings and do not account for the spatial-polarization coupling induced by antenna rotation in RIS-empowered SR systems. 

For RIS-empowered SR systems, the adjustable radiation patterns of RAs at BS can directly enhance the desired links and reduce interference. By rotating the BS antennas, their radiation patterns can be steered toward desired propagation directions, including those toward the DP-RIS and the SR user. Such directional gain enhancement helps alleviate the severe double-fading effect in the RIS-assisted secondary link, while reducing radiation leakage toward coexisting non-SR users. However, realizing these benefits raises two design issues. First, the effective gain of the RIS-assisted secondary link depends on both antenna-boresight alignment and polarization matching. Since antenna rotation changes the radiation direction and the local polarization bases at the same time, the RA rotations need to be jointly designed with the transceiver polarization states and DP-RIS phase shifts. Second, independently optimizing a continuous rotation matrix for each antenna incurs considerable computational and rotation-control complexity, motivating the development of low-complexity RA designs for practical deployment.\looseness-1

Motivated by the above challenges, this paper investigates a DP-RIS-empowered SR system where the BS is equipped with polarization-reconfigurable RAs. 
The proposed design jointly exploits antenna orientations, transceiver polarization states, and DP-RIS phase shifts to enhance the primary and secondary transmissions while protecting coexisting non-SR users. 
Different from conventional RA designs that mainly focus on spatial-domain reconfiguration, we explicitly account for the changes in local polarization bases caused by antenna rotation and develop a joint spatial-polarization design framework.
To improve practical implementability, we further develop two low-complexity RA designs. The first is a subarray-wise shared rotation design, where antennas in the same subarray share a common rotation matrix to reduce rotation control complexity. The second is a discrete RA rotation codebook design, where each antenna selects its rotation matrix from a finite codebook to avoid continuous rotation optimization.
The main contributions of this paper are summarized as follows.
\begin{itemize}
\item We investigate a DP-RIS-empowered SR system where the BS is equipped with polarization-reconfigurable RAs. 
In this system, RAs provide orientation reconfigurability for spatial-domain gain adjustment, polarforming enables transmit polarization reconfiguration, and the DP-RIS assists the secondary transmission through polarization-dependent passive reflection.
\item We formulate a transmit power minimization problem by jointly optimizing digital beamforming, RA rotation matrices, transceiver polarization states, and DP-RIS phase shifts under the primary and secondary rate requirements, interference temperature constraints, and hardware constraints. To tackle the resulting non-convex problem, we develop an alternating optimization (AO)-based algorithm that combines semidefinite programming (SDP), difference-of-convex (DC) programming, and Riemannian conjugate gradient (RCG) methods.
\item To reduce hardware and computational complexity, we develop two low-complexity RA designs: a subarray-wise shared rotation design and a discrete RA rotation codebook design. The former lets antennas in the same subarray share a common rotation matrix, while the latter selects the rotation matrix from a finite codebook constructed according to the SR-user and DP-RIS directions, thereby reducing rotation control and optimization complexity.
\item Simulation results demonstrate that the proposed joint spatial-polarization design significantly reduces the required transmit power compared with benchmark schemes using fixed antenna orientations or fixed polarization states. The results also show that antenna rotation effectively exploits high directivity gains, while polarforming mitigates polarization mismatch and improves the SR system performance. Moreover, the proposed low-complexity RA designs achieve comparable performance with lower hardware and computational complexity.
\end{itemize} 

The rest of the paper is organized as follows: 
Section \uppercase\expandafter{\romannumeral2} introduces the system model of the proposed DP-RIS-empowered SR system with polarization-reconfigurable RAs. 
Section \uppercase\expandafter{\romannumeral3} presents a toy example to illustrate the insightst of RA design. 
Section \uppercase\expandafter{\romannumeral4} formulates the transmit power minimization problem. 
Section \uppercase\expandafter{\romannumeral5} develops the AO algorithm based on SDP, DC programming, and RCG methods. Section \uppercase\expandafter{\romannumeral6} introduces the proposed low-complexity rotation designs. 
Simulation results are presented in Section \uppercase\expandafter{\romannumeral7}, and Section \uppercase\expandafter{\romannumeral8} concludes the paper.

Notations: Lowercase and uppercase bold letters represent vectors and matrices, respectively. The conjugate, transpose and conjugate transpose of matrix $\bf A $ are denoted by ${\bf A}^* $, ${\bf A}^T $ and ${\bf A}^H $, respectively. $\mathbb{C}^{M \times N}$ and $\mathbb{R}^{M \times N}$ denote the spaces of $M \times N$ complex and real matrices, respectively. $\mathbf{A} \succeq \mathbf{0}$ indicates that $\mathbf{A}$ is a positive semi-definite (PSD) matrix. $\mathrm{Tr}(\mathbf{A})$ and $\|\mathbf{A}\|_2$ denote the trace and spectral norm of matrix $\mathbf{A}$, respectively. For a vector $\mathbf{a}$, $\|\mathbf{a}\|$ denotes its Euclidean norm, and $|[\mathbf{a}]_i|$ denotes the modulus of its $i$-th element. The symbol $\odot$ represents the Hadamard product, and $\mathfrak{R}\{\cdot\}$ denotes the real part of a complex value. Finally, $\mathcal{CN}(\mu, \sigma^2)$ represents the circularly symmetric complex Gaussian (CSCG) distribution with mean $\mu$ and variance $\sigma^2$.
\section{System Model}
\begin{figure}[t!]
  \centering
  \includegraphics[width=0.45\textwidth]{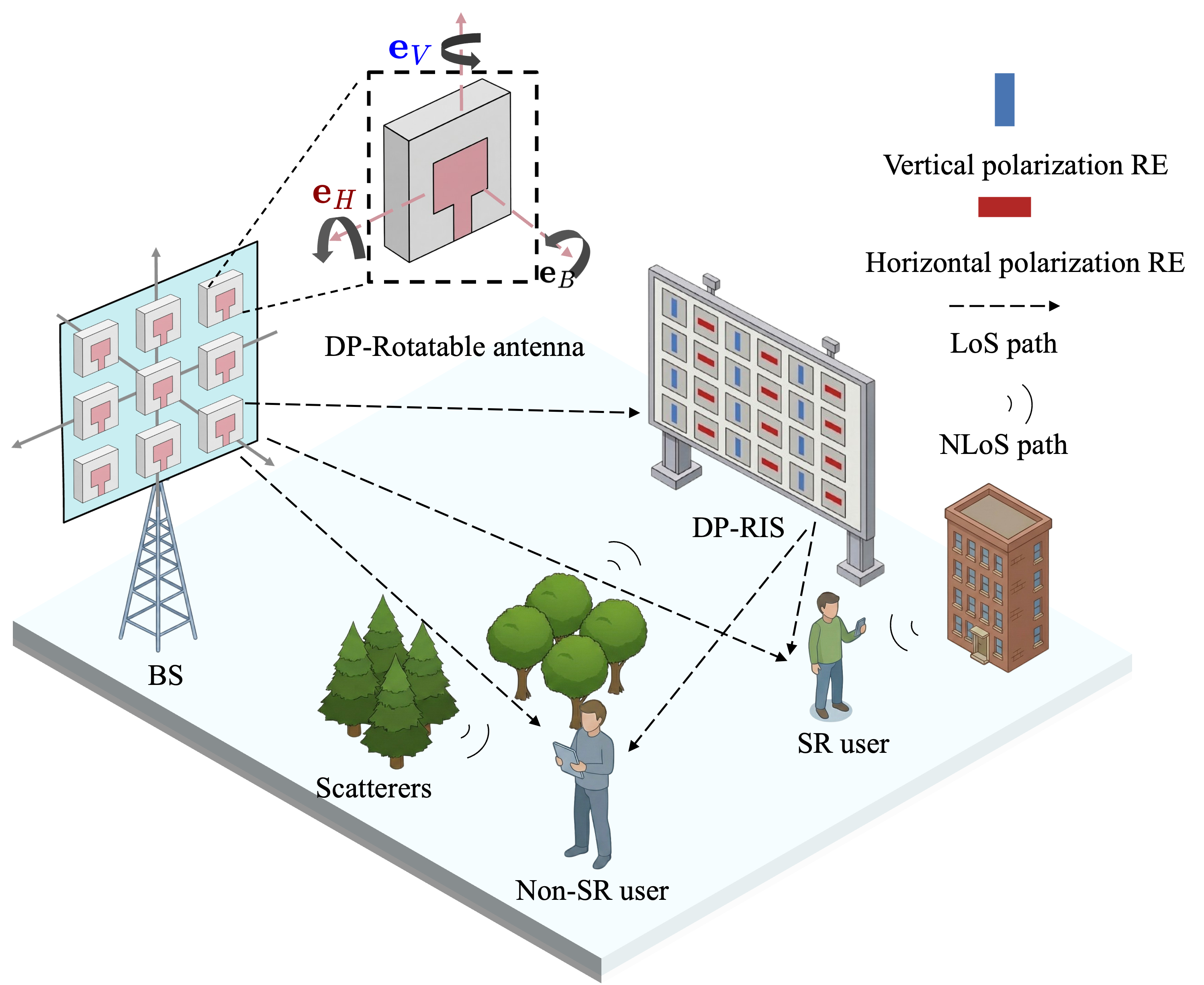}
  \caption{The DP-RIS-empowered SR system with RAs}
  \label{figure}
\end{figure}

As illustrated in Fig. \ref{figure}, we consider a DP-RIS-empowered SR system comprising one SR user and $K$ non-SR users, where the BS and the DP-RIS are equipped with $M$ RAs and $N$ reflecting elements (REs) arranged in uniform planar arrays (UPAs), respectively. Specifically, each antenna at the BS integrates co-located horizontal (H) and vertical (V) polarized ports. The DP-RIS consists of V- and H-polarized REs arranged in a column-wise interleaved fashion \cite{10785256}.
\subsection{Rotation, Polarization, and DP-RIS Configuration}
Without loss of generality, we establish a global Cartesian coordinate system where the initial boresight, H-polarized, and V-polarized directions are aligned with the x-, y-, and z-axes, denoted by $\mathbf{e}_B = [1, 0, 0]^T$, $\mathbf{e}_H = [0, 1, 0]^T$, and $\mathbf{e}_V = [0, 0, 1]^T$, respectively. Furthermore, let $\mathbf{R}_m\!=\![\mathbf{r}_{m,1},\!\mathbf{r}_{m,2},\!\mathbf{r}_{m,3}]\! \in \!\mathbb{R}^{3 \times 3}$ denote the rotation matrix for the $m$-th antenna, which is an orthogonal matrix satisfying $\mathbf{R}_m^T \mathbf{R}_m = \mathbf{I}_3$ and $\det(\mathbf{R}_m) = 1$. Consequently, the rotated boresight, H-polarized, and V-polarized directions are given by the columns $ \mathbf{r}_{m,1}$, $\mathbf{r}_{m,2}$, and $\mathbf{r}_{m,3}$, respectively. It is worth noting that the above rotation-matrix representation and the used roll, pitch, and yaw angle representation \cite{6DMA} both describe the same three-dimensional (3D) antenna rotation.

Beyond spatial rotation, the BS employs polarforming to reconfigure the polarization state by adjusting complex weights at the dual-polarized ports via power splitters and phase shifters~\cite{PRA,11347516}. Specifically, the transmit polarization state of the $m$-th BS antenna is characterized by $\mathbf{v}_m = [\rho_{m,H} e^{j \phi_{m,H}}, \rho_{m,V} e^{j \phi_{m,V}}]^T \in \mathbb{C}^{2 \times 1}$, satisfying $\rho_{m,H}^2 + \rho_{m,V}^2 = 1$ and $\phi_{m,H}, \phi_{m,V} \in [0, 2\pi)$. For the SR user, we consider a normalized receive polarization vector $\mathbf{u}_0 \in \mathbb{C}^{2 \times 1}$ to characterize its adjustable receive polarization state for polarization matching. In contrast, the $K$ non-SR users are equipped with fixed polarization antennas, whose polarization states are denoted by $\mathbf{u}_k$ for all $k=1,\dots,K$.

For the DP-RIS, let $\mathbf{q}_H, \mathbf{q}_V \in \mathbb{R}^{3 \times 1}$ denote the reference H and V polarization directions in the global coordinate system. Accordingly, the physical polarization direction of the $n$-th RE is given by $\mathbf{q}_n \in \{\mathbf{q}_H, \mathbf{q}_V\}$. Let $\boldsymbol{\theta} = [\theta_1, \theta_2, \dots, \theta_N]^T$ be the phase shift vector comprising independent phase shifts $\theta_n \in [0, 2\pi)$, and the corresponding phase shift matrix is given by $\mathbf{\Theta} = \text{diag}(e^{j\theta_1}, e^{j\theta_2}, \dots, e^{j\theta_N}) \in \mathbb{C}^{N \times N}$.

\subsection{Channel Model}
We adopt a polarization channel model comprising one line-of-sight (LoS) path and multiple non-LoS (NLoS) paths~\cite{he2022polarization}. The effective transmitted polarization vector for the $m$-th BS antenna is given by $\mathbf{e}_{t,m} = [\mathbf{r}_{m,2}, \mathbf{r}_{m,3}] \mathbf{v}_m \in \mathbb{C}^{3 \times 1}$. Let $d_{n,m,l}^{BR}$ denote the total propagation distance for the $l$-th path, with $\mathbf{k}_{n,m,l}^{BR,tx}, \mathbf{k}_{n,m,l}^{BR,rx} \in \mathbb{R}^{3 \times 1}$ representing the corresponding normalized departure and arrival direction vectors, respectively. Specifically, for the LoS path ($l=0$), both vectors point directly from the $m$-th BS antenna to the $n$-th DP-RIS element. For the NLoS paths ($l \ge 1$), $\mathbf{k}_{n,m,l}^{BR,tx}$ points from the BS antenna to the scatterer, while $\mathbf{k}_{n,m,l}^{BR,rx}$ points from the scatterer to the DP-RIS. After rotation, the $m$-th antenna boresight direction becomes $\mathbf{r}_{m,1}$, and the large-scale fading coefficient for the $l$-th path is formulated as 
\begin{align*}
\beta_{n,m,l}^{BR} \approx \frac{A}{4\pi(d_{n,m,l}^{BR})^{\alpha_l^{BR}}} G_0 \left( \max\{0,\mathbf{r}_{m,1}^T \mathbf{k}_{n,m,l}^{BR,tx}\} \right)^{2p},\tag{1} \label{1}
\end{align*}
where $A$ denotes the physical aperture size of the antenna, $\alpha_l^{BR}$ is the path loss exponent specific to the $l$-th path, $G_0\!=2(2p+1)$ represents the peak directional gain, and $p$ is the directivity factor~\cite{11427014}.
For the small-scale fading, let $\mathbf{Z}_{n,m,l}^{BR,tx}, \mathbf{Z}_{n,m,l}^{BR,rx} \in \mathbb{R}^{3 \times 2}$ denote the orthonormal bases orthogonal to the propagation directions $\mathbf{k}_{n,m,l}^{BR,tx}$ and $\mathbf{k}_{n,m,l}^{BR,rx}$, respectively. 
Furthermore, let $\mathbf{M}_{n,m,l}^{BR} \in \mathbb{C}^{2 \times 2}$ denote the polarization coupling matrix. Assuming ideal polarization preservation for the LoS path, we set $\mathbf{M}_{n,m,0}^{BR} = \mathbf{I}_2$. For the NLoS paths, the coupling matrices are modeled as
\begin{align*}
\mathbf{M}_{n,m,l}^{BR} = \begin{bmatrix} \sqrt{1-\chi} e^{j\phi_{n,m,l}^{HH}} & \sqrt{\chi} e^{j\phi_{n,m,l}^{HV}} \\ \sqrt{\chi} e^{j\phi_{n,m,l}^{VH}} & \sqrt{1-\chi} e^{j\phi_{n,m,l}^{VV}} \end{bmatrix},l \ge 1,\tag{2} \label{2}
\end{align*}
where $\chi \in (0,1]$ denotes the cross-polarization leakage power fraction, and $\phi_{n,m,l}^{ab} \in [0, 2\pi)$ for $a,b \in \{H,V\}$ represent the independent random phase shifts introduced by the scatterer. Consequently, the channel coefficient $g_{n,m}$, projected onto the polarization $\mathbf{q}_n$ of the $n$-th RE, is formulated as~\cite{he2022polarization}
\begin{align*}
g_{n,m} = \sum_{l=0}^{L_{BR}} \sqrt{\beta_{n,m,l}^{BR}} e^{-j\frac{2\pi}{\lambda} d_{n,m,l}^{BR}} \mathbf{q}_n^T  \mathbf{B}_{n,m,l}^{BR}\mathbf{e}_{t,m},\tag{3} \label{3}
\end{align*}
where ${L_{BR}}$ denotes the number of paths. $\mathbf{B}_{n,m,l}^{BR}=\mathbf{Z}_{n,m,l}^{BR,rx} \mathbf{M}_{n,m,l}^{BR} (\mathbf{Z}_{n,m,l}^{BR,tx})^T$ accounts for the channel depolarization effect. By collecting all the channel coefficients $g_{n,m}$, the channel matrix from the BS to the DP-RIS is $\mathbf{G} = [g_{n,m}]_{n,m} \in \mathbb{C}^{N \times M}$. 

Following a similar modeling procedure, we establish the direct channels and the reflected channels. To facilitate a unified representation, let $i \in \{0, 1, \dots, K\}$ denote the user index, where $i=0$ corresponds to the SR user and $i \ge 1$ corresponds to the $K$ non-SR users. Let $\mathbf{E}_{rx} = [\mathbf{e}_H, \mathbf{e}_V] \in \mathbb{R}^{3 \times 2}$ denote the reference polarization basis at the receivers. The channel coefficient $h_{i,m}$ from the $m$-th BS antenna to user $i$, and the reflected channel coefficient $f_{i,n}$ from the $n$-th RE to user $i$, are respectively formulated as
\begin{align*}
h_{i,m} &= \sum_{l=0}^{L_{BU, i}} \sqrt{\beta_{i,m,l}^{BU}} e^{-j\frac{2\pi}{\lambda} d_{i,m,l}^{BU}} \mathbf{u}_i^H \mathbf{E}_{rx}^T \mathbf{B}_{i,m,l}^{BU} \mathbf{e}_{t,m},\tag{4} \label{4}\\ 
f_{i,n} &= \sum_{l=0}^{L_{RU, i}} \sqrt{\beta_{i,n,l}^{RU}} e^{-j\frac{2\pi}{\lambda} d_{i,n,l}^{RU}} \mathbf{u}_i^H \mathbf{E}_{rx}^T \mathbf{B}_{i,n,l}^{RU} \mathbf{q}_n,\tag{5} \label{5}
\end{align*}
where $\{L_{BU, i},L_{RU, i}\}$, $\{\beta_{i,m,l}^{BU},\beta_{i,n,l}^{RU}\}$, and $\{d_{i,m,l}^{BU},d_{i,n,l}^{RU}\}$ represent the number of paths, large-scale fading coefficients, and propagation distances for the respective links, while $\{\mathbf{B}_{i,m,l}^{BU}, \mathbf{B}_{i,n,l}^{RU}\}$ model the channel depolarization effect. Note that $\beta_{i,n,l}^{RU}$ is modeled assuming omnidirectional antennas. Finally, the individual coefficients are assembled into the channel vectors $\mathbf{h}_i = [h_{i,1}, \dots, h_{i,M}]^T \in \mathbb{C}^{M \times 1}$ and $\mathbf{f}_i = [f_{i,1}, \dots, f_{i,N}]^T \in \mathbb{C}^{N \times 1}$.
\subsection{Transmission Model}
Let $s(t)$ denote the primary symbol for the SR user ($i=0$) at index $t$, satisfying $\mathbb{E}[|s(t)|^2] = 1$, which is transmitted via the BS beamforming vector $\mathbf{w} \in \mathbb{C}^{M \times 1}$. 
Additionally, the DP-RIS modulates its BPSK symbol $c \in \{+1, -1\}$ by periodically adjusting phase shifts of all REs, modeled as $\mathbf{\Theta}(c) = c \mathbf{\Theta}$. We assume the symbol period ratio between the secondary and primary systems satisfies $N_T = T_c / T_s \gg 1$. During one DP-RIS symbol period $T_c$, the BPSK symbol $c$ remains constant, acting as a slow-varying channel. After receive polarization processing, the received signal for user $i \in \{0, 1, \dots, K\}$ at index $t \in \{1, 2, \dots, N_T\}$ is given by\looseness-1
\begin{align*}
y_i(t) = \left(\mathbf{h}_i^H + c \mathbf{f}_i^H \mathbf{\Theta} \mathbf{G} \right) \mathbf{w} s(t) + n_i(t),\tag{6} \label{6}
\end{align*}
where $n_i(t) \sim \mathcal{CN}(0, \sigma^2)$ denotes the additive white Gaussian noise.
Thus, the achievable rate of the primary system is expressed as  \looseness-1
\begin{align*}
R_s(c)= \log_2\left( 1 + \frac{|\left(\mathbf{h}_0^H + c \mathbf{f}_0^H \mathbf{\Theta} \mathbf{G}\right) \mathbf{w}|^2}{\sigma^2} \right).\tag{7} \label{7}
\end{align*}
Following the successful decoding of $s(t)$, the SR user employs successive interference cancellation (SIC) to remove the primary signal component of the direct link from $y_0(t)$. By applying maximal-ratio combining (MRC) over the $N_T$ residual samples, a processing gain of $N_T$ is achieved, yielding the achievable rate of signal $c$ as
\begin{align*}
R_c =\frac{1}{N_T}\log_2\left(1+ \frac{N_T |\mathbf{f}_0^H \mathbf{\Theta} \mathbf{G} \mathbf{w}|^2}{\sigma^2}\right).\tag{8} \label{8}
\end{align*}
Moreover, the non-SR users suffer from the interference originating from both the direct BS transmission and the DP-RIS reflection. Assuming the BPSK symbol $c \in \{+1, -1\}$ is equally probable, the average interference power at the $k$-th non-SR user can be expressed as
\begin{align*}
\bar{I}_k \!=\! \mathbb{E}_c \left[ |\left(\mathbf{h}_k^H \!\!+ \!c \mathbf{f}_k^H \mathbf{\Theta} \mathbf{G}\right) \mathbf{w}|^2 \right] 
\!= \!|\mathbf{h}_k^H \mathbf{w}|^2 \!+ \!|\mathbf{f}_k^H \mathbf{\Theta} \mathbf{G} \mathbf{w}|^2.\tag{9} \label{9}
\end{align*}
To protect the communication quality of the non-SR users, this average interference power must be strictly bounded by a predefined threshold $I_{\text{limit}}$.
\section{Toy Example: Insights on RA Design}
\begin{figure}[t]
    \centering
    \subfloat[Isotropic antenna]{
        \includegraphics[width=0.48\columnwidth]{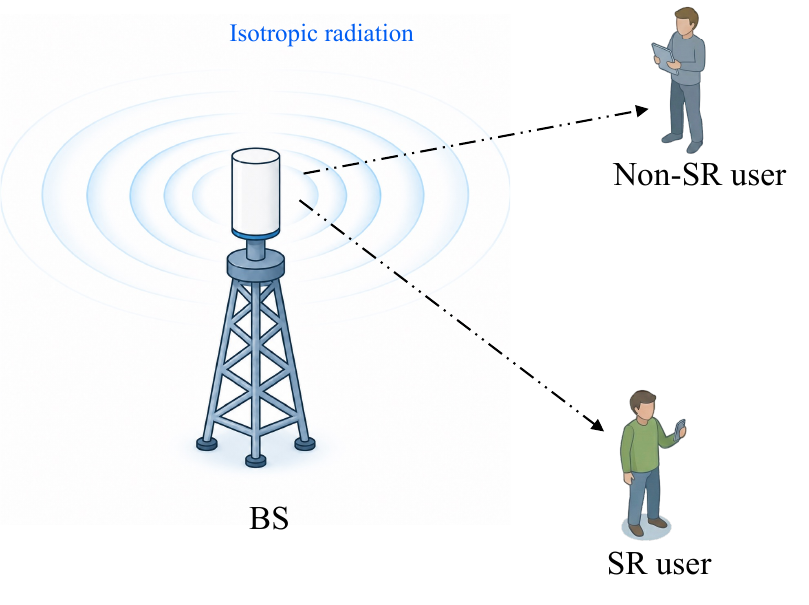}
        \label{fig:toy_omni}
    }
    \subfloat[Rotatable antenna]{
        \includegraphics[width=0.48\columnwidth]{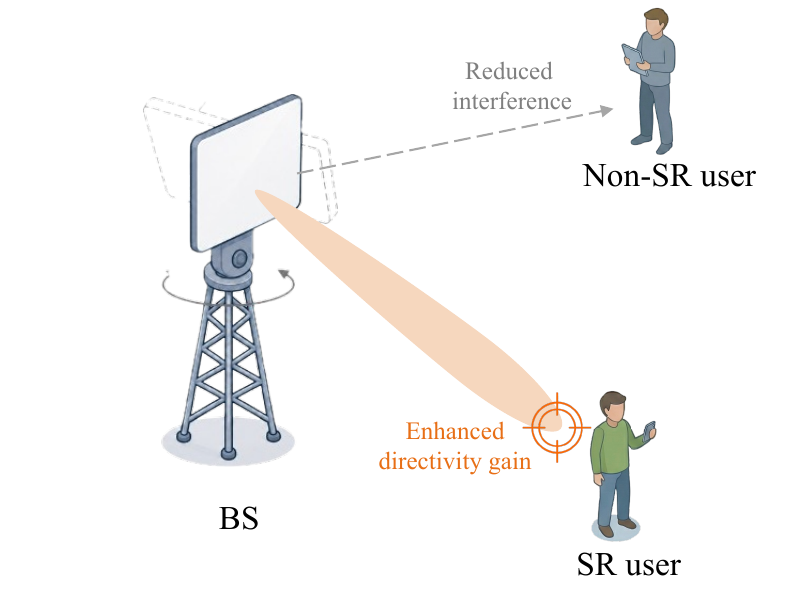}
        \label{fig:toy_ra}
    }
    \caption{RA-enabled gain enhancement and interference reduction}
    \label{fig:toy_example}
\end{figure}

To provide intuition for the spatial benefit of rotatable antennas (RAs) and the subsequent low-complexity rotation designs, we consider a simple LoS setup with one SR user and one non-SR user. For clarity, the RIS-reflected link and multipath components are omitted in this section. As illustrated in Fig.~\ref{fig:toy_example}, an isotropic antenna radiates energy uniformly in all directions, whereas a rotatable antenna can steer its radiation pattern by adjusting its boresight direction.

First, we illustrate the capability of RAs to simultaneously enhance the desired signal and suppress interference. In the considered setup, the large-scale channel gain is determined by the antenna directional gain, which strongly depends on the alignment between the antenna boresight direction and the signal departure direction. According to the RA channel model in~\eqref{1}, the directional gain toward direction $\mathbf{k}$ is given by
$
G_0\left(\max\{0,\mathbf{r}_{m,1}^{T}\mathbf{k}\}\right)^{2p}
$, 
where $\mathbf{r}_{m,1}$ denotes the rotated boresight direction and $p$ is the directivity factor.
As illustrated in Fig.~\ref{fig:toy_example}, an isotropic antenna radiates energy uniformly in all directions, and therefore provides no capability to shape spatial radiation or suppress interference toward undesired directions without beamforming. In contrast, a rotatable antenna can steer its radiation pattern by adjusting its boresight direction, enabling spatial focusing toward desired directions and reduced leakage toward undesired directions.

\begin{figure}[t!]
  \centering
  \includegraphics[width=0.48\textwidth]{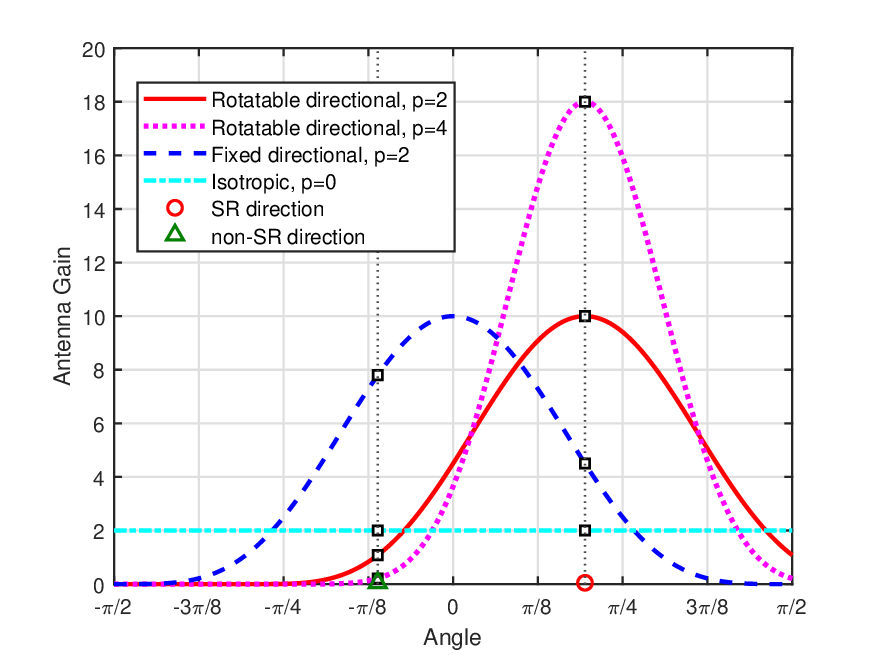}
  \caption{Antenna gain versus angle}
  \label{figure_angle}
\end{figure}

Fig.~\ref{figure_angle} further quantifies this effect by showing the antenna gain versus angle under different antenna configurations. In this example, the SR user and non-SR user are located at angular directions of $35^\circ$ and $-20^\circ$, respectively. The fixed directional antenna has a preset boresight direction that does not align with the SR user, while the RA dynamically steers its boresight toward the SR-user direction. As a result, the RA achieves higher gain in the desired direction and lower gain in the non-SR direction. Moreover, a larger directivity factor $p$ further sharpens the angular selectivity, resulting in reduced radiation leakage toward undesired directions and making RAs a powerful physical tool for mitigating interference.

This observation also provides useful insights for low-complexity RA design. Since the spatial gain mainly depends on the alignment between the antenna boresight and a few dominant directions, useful rotation choices are usually associated with these directions rather than the entire continuous rotation space. Therefore, when the number of antennas is large, independently optimizing a continuous rotation matrix for each antenna may introduce considerable redundancy.
This motivates the development of low-complexity RA architectures in Section VI. On one hand, multiple antennas can share a common rotation matrix to reduce the number of rotation variables. On the other hand, the continuous rotation search can be replaced by a small codebook constructed from the main desired directions and their intermediate directions. In this way, the low-complexity designs reduce hardware and optimization complexity while preserving most of the spatial gain.\looseness-1

\section{Problem Formulation}
Although the toy example provides useful insights on RA design, the RIS-empowered SR system involves more tightly coupled design requirements. Specifically, the primary SR transmission and the RIS transmission are associated with different directions and may favor different RA orientations. In addition, antenna rotation changes not only the directional radiation gain but also the local polarization bases, which couples RA orientation with transceiver polarization states and DP-RIS reflection. Therefore, in this paper, we propose a joint design method to minimize the total transmit power by optimizing the rotation matrices $\mathbf{R} = \{\mathbf{R}_m\}_{m=1}^M$, the transmit polarization vectors $\mathbf{V} = \{\mathbf{v}_m\}_{m=1}^M$, the digital beamforming vector $\mathbf{w}$, the DP-RIS phase shift matrix $\mathbf{\Theta}$, and the receive polarization vector $\mathbf{u}_0$ at the SR user. The optimization is subject to the rate requirements of both primary and secondary transmissions, the interference temperature constraints for non-SR users, and practical hardware constraints, as formulated below.\looseness-1
\begin{align*}
\mathbf{P}1\min_{ \mathbf{R}, \mathbf{V}, \mathbf{\Theta},\mathbf{u}_0,\mathbf{w}} \quad & \|\mathbf{w}\|^2 \tag{10} \label{10}\\
\text{s.t.} \quad & R_s(c) \geq \bar{R}_s,\forall c, \tag{10a} \label{10a}\\
& R_c \geq \bar{R}_c, \tag{10b} \label{10b} \\
& |\mathbf{h}_k^H \mathbf{w}|^2 + |\mathbf{f}_k^H \mathbf{\Theta} \mathbf{G} \mathbf{w}|^2  \leq I_{\text{limit}}, \forall k,  \tag{10c} \label{10c}\\
& \mathbf{R}_m \mathbf{R}_m^T={\bf I}_3,\det(\mathbf{R}_m)=1, \forall m,\tag{10d} \label{10d}\\ 
&0\le \arccos(\mathbf{r}_{m,1}^T{\bf e}_B) \le \theta_{\max}, \forall m, \tag{10e} \label{10e}\\
&\|\mathbf{v}_m\|^2 = 1,\forall m, \tag{10f} \label{10f}\\
&\|\mathbf{u}_0\|^2 = 1, \tag{10g} \label{10g}\\
&|\mathbf{\Theta}(i,i)|=1,\forall i \tag{10h} \label{10h},
\end{align*}
where \eqref{10a} ensures that the primary transmission rate achieves the threshold $\bar{R}_s$ for both states of the DP-RIS symbol $c$, while \eqref{10b} enforces the minimum target rate $\bar{R}_c$ for the secondary transmission. Furthermore, constraint \eqref{10c} guarantees that the average interference power leaked to each non-SR user remains below $I_{\text{limit}}$. From the hardware perspective, \eqref{10d} constrains the rotation matrices to belong to the special orthogonal group $\mathrm{SO}(3)$, while \eqref{10e} limits the rotation relative to the initial boresight $\mathbf{e}_B$ within $\theta_{\max}$. Constraints \eqref{10f} and \eqref{10g} represent the unit-norm constraints on the transceiver polarization states, respectively. Finally, \eqref{10h} represents the unit-modulus constraint for the passive DP-RIS. \looseness-1
\section{Proposed Solutions}
Due to the tightly coupled variables and hardware constraints, problem $\mathbf{P}1$ is highly non-convex. We propose an AO framework to update each variable iteratively. Specifically, the digital beamforming $\mathbf{w}$ is optimized via SDP and DC programming. For the remaining variables $\mathbf{R}$, $\{\mathbf{V}, \mathbf{u}_0\}$, and $\mathbf{\Theta}$, we employ the RCG method to optimize them directly on their respective manifolds: the $\mathrm{SO}(3)$ group, complex unit sphere, and complex circle manifold.\looseness-1
\subsection{Optimization of the Digital Beamforming at the BS}
Given $\mathbf{R}$, $\{\mathbf{V}, \mathbf{u}_0\}$, and $\mathbf{\Theta}$, we optimize $\mathbf{w}$ by introducing a positive semidefinite (PSD) matrix $\mathbf{D} \triangleq \mathbf{w}\mathbf{w}^H$ to tackle the non-convex quadratic terms in \eqref{10a}-\eqref{10c}. By defining the equivalent channel vector as $\mathbf{h}_s^H(c) \triangleq \mathbf{h}_0^H + c\mathbf{f}_0^H\mathbf{\Theta}\mathbf{G}$, the optimization subproblem with respect to $\mathbf{D}$ is formulated as:
\begin{align*}
\mathbf{P}2\;\min_{ \mathbf{D} \succeq \mathbf{0}} \; & \text{Tr}(\mathbf{D}) \tag{11} \label{11}\\
\text{s.t.} \;& \text{Tr}\left( \mathbf{h}_s(c) \mathbf{h}_s^H(c) \mathbf{D} \right) \ge \Gamma_s, \forall c, \tag{11a} \label{11a}\\
& \text{Tr}\left( \mathbf{G}^H \mathbf{\Theta}^H \mathbf{f}_0 \mathbf{f}_0^H \mathbf{\Theta} \mathbf{G} \mathbf{D} \right) \ge \Gamma_c, \tag{11b} \label{11b} \\
& \text{Tr}\left( \!\left(\mathbf{h}_k \mathbf{h}_k^H \!\!+\! \mathbf{G}^H \!\mathbf{\Theta}^H \!\mathbf{f}_k \mathbf{f}_k^H \!\mathbf{\Theta} \mathbf{G}\right)\! \mathbf{D}\! \right) \!\le\! I_{\text{limit}}, \!\forall k,  \tag{11c} \label{11c}\\
& \text{rank}(\mathbf{D})=1, \tag{11d} \label{11d}
\end{align*}
where $\Gamma_s = (2^{\bar{R}_s} - 1)\sigma^2$ and $\Gamma_c = \frac{2^{N_T \bar{R}_c} - 1}{N_T} \sigma^2$. Problem $\mathbf{P}2$ remains non-convex due to the rank-one constraint \eqref{11d}. To tackle this, we introduce the following equivalence, which holds for any PSD matrix $\mathbf{D} \succeq \mathbf{0}$
\begin{align*}
\text{rank}(\mathbf{D}) = 1 \iff \text{Tr}(\mathbf{D}) - \lambda_{\max}(\mathbf{D}) = 0, \tag{12} \label{12}
\end{align*}
where $\lambda_{\max}(\mathbf{D})$ denotes the maximum eigenvalue of $\mathbf{D}$. By incorporating $\text{Tr}(\mathbf{D}) - \lambda_{\max}(\mathbf{D})$ as a penalty term into the objective function with a factor $\eta > 0$, the original problem becomes a DC program. Specifically, to handle the concave term $-\lambda_{\max}(\mathbf{D})$, we linearize it via the first-order Taylor expansion at the previous iterate $\mathbf{D}^{(t-1)}$, yielding the convex upper bound $\hat{f}(\mathbf{D}) = -\lambda_{\max}(\mathbf{D}^{(t-1)}) - \Re\{\text{Tr}( \mathbf{d}^{(t-1)}(\mathbf{d}^{(t-1)})^H (\mathbf{D} - \mathbf{D}^{(t-1)}) )\}$, where $\mathbf{d}^{(t-1)}$ is the principal eigenvector of $\mathbf{D}^{(t-1)}$. By removing the constant terms independent of $\mathbf{D}$, the subproblem at iteration $t$ is reformulated as
\begin{align*}
\mathbf{P}2.1 \;\min_{\mathbf{D} \succeq \mathbf{0}} \; & (1+\eta)\text{Tr}(\mathbf{D}) - \eta \Re\{\text{Tr}(\mathbf{d}^{(t-1)}(\mathbf{d}^{(t-1)})^H \mathbf{D})\} \notag \\
\text{s.t.} \; & \eqref{11a}, \eqref{11b}, \eqref{11c}.
\end{align*}
The $\mathbf{P}2.1$ is convex and can be solved by CVX tools \cite{grant2008cvx}.
\subsection{Optimization of the Rotation Matrices at the BS}
Given $\mathbf{w}$, $\{\mathbf{V}, \mathbf{u}_0\}$, and $\mathbf{\Theta}$, the objective function of $\mathbf{P}1$ is independent of $\mathbf{R}$. To find a feasible solution that satisfies all constraints, we adopt a max-min approach based on the normalized constraint margins. Specifically, we define $\beta_{s,1}= \frac{|(\mathbf{h}_0^H + \mathbf{f}_0^H \mathbf{\Theta} \mathbf{G})\mathbf{w}|^2}{\Gamma_s}$, $\beta_{s,2}= \frac{|(\mathbf{h}_0^H - \mathbf{f}_0^H \mathbf{\Theta} \mathbf{G})\mathbf{w}|^2}{\Gamma_s}$, $\beta_{c}= \frac{|\mathbf{f}_0^H \mathbf{\Theta} \mathbf{G} \mathbf{w}|^2}{\Gamma_c}$ and $\beta_{k,\mathrm{int}} = \frac{|\mathbf{h}_k^H \mathbf{w}|^2 + |\mathbf{f}_k^H \mathbf{\Theta} \mathbf{G} \mathbf{w}|^2}{I_{\text{limit}}}$. 
Here, $\beta_{s,i} \ge 1$ and $\beta_c \ge 1$ indicate that the corresponding rate requirements are satisfied, whereas $\beta_{k,\mathrm{int}} \le 1$ ensures that the interference temperature constraint is satisfied. To enable the use of the RCG method, we reformulate the optimization of $\mathbf{R}$ as an unconstrained problem on the product manifold $\mathcal{M}_R=\mathrm{SO}(3)^M$. 
By integrating the worst-case margin with quadratic penalties for constraint violations, the objective function is given by
\begin{align*}
\mathbf{P}3\;\min_{\mathbf{R}\in\mathcal{M}_R}\; f_R
&= f_1 + \lambda_2 f_2 + \lambda_3 f_3 + \lambda_4 f_4 + \lambda_5 f_5, \tag{13} \label{13}
\end{align*}
where $f_1 = -\min\{\beta_{s,1},\beta_{s,2},\beta_c\}$ aims to maximize the minimum margin. The penalty terms $f_2 = \sum_{i=1}^2\max(0,1-\beta_{s,i})^2$, $f_3 = \max(0,1-\beta_c)^2$, $f_4 = \sum_{k=1}^{K}\max(0,\beta_{k,\mathrm{int}}-1)^2$, and $f_5 = \sum_{m=1}^{M}\max(0,\cos\theta_{\max}-\mathbf{r}_{m,1}^T\mathbf{e}_B)^2$ enforce the constraints for the primary system, secondary system, interference temperature, and rotation range, respectively, with $\lambda_j > 0$ being the penalty parameter associated with each $f_j$.

Since the objective function $f_R$ remains non-smooth due to the $\min(\cdot)$ and $\max(\cdot)$ operators, we further introduce a smooth approximation. Specifically, $\min\{\beta_{s,1},\beta_{s,2},\beta_c\}$ and $\max(0,x)$ are approximated by the log-sum-exp and softplus functions, respectively~\cite{boyd2004convex}. The resulting smooth function is  
\begin{align*}
\mathbf{P}3.1\;\min_{\mathbf{R}\in\mathcal{M}_R}\; \tilde f_R
&= \tilde f_1 + \lambda_2 \tilde f_2 + \lambda_3 \tilde f_3 + \lambda_4 \tilde f_4 + \lambda_5 \tilde f_5,\tag{14} \label{14}
\end{align*}
where $\tilde f_1= \frac{1}{\mu}\ln\!\left(e^{-\mu\beta_{s,1}} + e^{-\mu\beta_{s,2}} + e^{-\mu\beta_c}\right)$, $\tilde f_2 = \sum_{i=1}^2 S_+(1-\beta_{s,i}, \alpha_2)^2$, $\tilde f_3= S_+(1-\beta_c,\alpha_3)^2$, $\tilde f_4= \sum_{k=1}^{K} S_+(\beta_{k,\mathrm{int}}-1,\alpha_4)^2$ and $\tilde f_5= \sum_{m=1}^{M} S_+(\cos\theta_{\max}-\mathbf{r}_{m,1}^T\mathbf{e}_B,\alpha_5)^2$. Here, $S_+(x,\alpha)=\frac{1}{\alpha}\ln(1+e^{\alpha x})$ denotes the softplus function, while $\mu>0$ and $\alpha_j>0$ are the corresponding smoothing parameters. The smooth unconstrained problem on $\mathcal{M}_R = \mathrm{SO}(3)^M$ is then  solved using the RCG method.

Let \(\mathbf{T}_m=\nabla_{\mathbf{R}_m}\tilde f_R\) denote the Euclidean gradient of the smooth objective \(\tilde f_R\) with respect to \(\mathbf{R}_m\). Since \(\tilde f_R\) is composed of the effective channel terms 
\((\mathbf h_i^H+\mathbf f_i^H\boldsymbol{\Theta}\mathbf G)\mathbf w\), 
\(\mathbf f_i^H\boldsymbol{\Theta}\mathbf G\mathbf w\), 
and the rotation-range penalty, \(\mathbf T_m\) can be obtained by applying the chain rule and matrix calculus to these matrix-vector operations.
Since \(\mathbf R_m\in\mathrm{SO}(3)\), the Euclidean gradient \(\mathbf T_m\) is projected onto the tangent space of \(\mathrm{SO}(3)\) at \(\mathbf R_m\). 
The Riemannian gradient is given by
\begin{align*}
\mathrm{grad}_{\mathbf R_m}\tilde f_R
=
\mathbf T_m-\mathbf R_m\mathrm{sym}\left(\mathbf R_m^T\mathbf T_m\right),
\tag{15} \label{15}
\end{align*}
where \(\mathrm{sym}(\mathbf A)=\frac{1}{2}(\mathbf A+\mathbf A^T)\).

At the \(t\)-th RCG iteration, the search direction is updated in the tangent space as
\begin{align*}
\boldsymbol{\xi}_m^{(t)}
=
-\mathrm{grad}_{\mathbf R_m}\tilde f_R\left(\mathbf R^{(t)}\right)
+
\beta^{(t)}
\mathcal T_{\mathbf R_m^{(t-1)}\rightarrow \mathbf R_m^{(t)}}
\left(\boldsymbol{\xi}_m^{(t-1)}\right),
\tag{16} \label{16}
\end{align*}
where \(\mathcal T_{\mathbf R_m^{(t-1)}\rightarrow \mathbf R_m^{(t)}}(\cdot)\) denotes the vector transport from \(T_{\mathbf R_m^{(t-1)}}\mathrm{SO}(3)\) to \(T_{\mathbf R_m^{(t)}}\mathrm{SO}(3)\), and \(\beta^{(t)}\) is the conjugate coefficient, e.g., computed by the Polak--Ribi\`ere rule. 
For \(t=0\), the search direction is initialized as
\begin{align*}
\boldsymbol{\xi}_m^{(0)}
=
-\mathrm{grad}_{\mathbf R_m}\tilde f_R\left(\mathbf R^{(0)}\right). \tag{17} \label{17}
\end{align*}

The step size \(\alpha^{(t)}\) is determined by the Armijo backtracking line search. 
Given \(\alpha^{(t)}\), the tentative point is
\begin{align*}
\mathbf Y_m^{(t)}
=
\mathbf R_m^{(t)}+\alpha^{(t)}\boldsymbol{\xi}_m^{(t)}.\tag{18} \label{18}
\end{align*}
Since \(\mathbf Y_m^{(t)}\) may not satisfy the orthogonality and determinant constraints of \(\mathrm{SO}(3)\), it is retracted back to the manifold. 
Using the polar retraction, let the singular value decomposition of \(\mathbf Y_m^{(t)}\) be
$
\mathbf Y_m^{(t)}=\mathbf A_m\boldsymbol{\Sigma}_m\mathbf B_m^T
$.
Then, the updated rotation matrix is obtained as
\begin{align*}
\mathbf R_m^{(t+1)}
=
\mathbf A_m
\mathrm{diag}\left(1,1,\det(\mathbf A_m\mathbf B_m^T)\right)
\mathbf B_m^T.
\tag{19} \label{19}
\end{align*}

\subsection{Optimization of the Transceiver Polarization Vectors}
With the other variables fixed, the subproblem with respect to the transmit polarization vectors $\mathbf{V}$ can be solved using the same RCG framework as the rotation matrices update. Due to the unit-norm constraint in \eqref{10f}, the feasible set of $\mathbf{V}$ is given by the product of complex unit spheres
\begin{align*}
\mathcal{M}_V=\left\{\mathbf{V}\,\middle|\, \|\mathbf{v}_m\|^2=1,\ \forall m\right\}.\tag{20} \label{20}
\end{align*}
By substituting $\mathbf{V}$ as the variable, the corresponding smooth unconstrained problem on $\mathcal{M}_V$ is formulated as
\begin{align*}
\mathbf{P}4\;\min_{\mathbf{V}\in\mathcal{M}_V}\; \tilde f_V
&= \tilde f_{1}+\lambda_2 \tilde f_{2}+\lambda_3 \tilde f_{3}+\lambda_4 \tilde f_{4},\tag{21} \label{21}
\end{align*}
where $\tilde f_1$ through $\tilde f_4$ maintain their previous definitions, while $\tilde f_5$ is omitted since it is independent of $\mathbf{V}$.  Let \(\mathbf g_m=\nabla_{\mathbf v_m^*}\tilde f_V\) denote the Euclidean gradient with respect to \(\mathbf v_m^*\). 
The Riemannian gradient is obtained by projecting $\mathbf g_m$ onto the tangent space of the complex unit sphere at $\mathbf v_m$, i.e.,
\begin{align*}
\mathrm{grad}_{\mathbf v_m}\tilde f_V
=
\left(\mathbf I_2-\mathbf v_m\mathbf v_m^H\right)\mathbf g_m,\quad \forall m.\tag{22} \label{22}
\end{align*}
After the tangent-space search direction is obtained by the RCG rule, the updated point is retracted back to \(\mathcal M_V\) by normalization,
\begin{align*}
\mathbf v_m^{(t+1)}
=
\frac{\mathbf v_m^{(t)}+\alpha^{(t)}\boldsymbol{\xi}_{v,m}^{(t)}}
{\left\|\mathbf v_m^{(t)}+\alpha^{(t)}\boldsymbol{\xi}_{v,m}^{(t)}\right\|},
\quad \forall m,\tag{23} \label{23}
\end{align*}
where \(\alpha^{(t)}\) is determined by the Armijo backtracking rule.

Similarly, the optimization of the receive polarization vector $\mathbf{u}_0$ is formulated on a single complex unit sphere, denoted as $\mathcal{M}_u = \{\mathbf{u}_0 \mid \|\mathbf{u}_0\|^2=1\}$. The corresponding smooth subproblem is given by
\begin{align*}
\mathbf{P}5\;\min_{\mathbf{u}_0\in\mathcal{M}_u}\; \tilde f_u
&= \tilde f_{1}+\lambda_2 \tilde f_{2}+\lambda_3 \tilde f_{3}+\lambda_4 \tilde f_{4}.\tag{24} \label{24}
\end{align*}
Since $\mathbf u_0$ is also constrained on a complex unit sphere, problem $\mathbf{P}5$ can be solved using the same RCG projection and normalization retraction as $\mathbf{P}4$, and the details are omitted for brevity.
\subsection{Optimization of the DP-RIS Phase Shift Matrix}
Following the RCG framework, the subproblem with respect to the DP-RIS phase shift matrix $\mathbf{\Theta}$ is formulated on the complex circle manifold, where the unit-modulus constraints are imposed on its diagonal entries. The feasible set is defined as \looseness-1
\begin{align*}
\mathcal{M}_{\Theta}=\left\{\mathbf{\Theta} \,\middle|\, \mathbf{\Theta}=\mathrm{diag}(e^{j\theta_1},\ldots,e^{j\theta_N})\right\},\tag{25} \label{25}
\end{align*}
which corresponds to the complex circle manifold. By treating $\tilde f_1$--$\tilde f_4$ as functions of $\mathbf{\Theta}$ with the other variables fixed, the smooth unconstrained subproblem with respect to $\mathbf{\Theta}$ can be written as
\begin{align*}
\mathbf{P}6\;\min_{\mathbf{\Theta}\in\mathcal{M}_{\Theta}}\; \tilde f_{\Theta}
= \tilde f_1+\lambda_2\tilde f_2+\lambda_3\tilde f_3+\lambda_4\tilde f_4.
\tag{26} \label{26}
\end{align*}

Let $\mathbf{G}_{\Theta}=\nabla_{\mathbf{\Theta}^*}\tilde f_{\Theta}$ denote the Euclidean gradient with respect to $\mathbf{\Theta}^*$. 
Since only the diagonal entries of $\mathbf{\Theta}$ are optimization variables, we define $\mathbf g_{\Theta}=\mathrm{diag}(\mathbf{G}_{\Theta})$ and $\boldsymbol{\vartheta}=\mathrm{diag}(\mathbf{\Theta})$. 
The Riemannian gradient is obtained by projecting $\mathbf g_{\Theta}$ onto the tangent space of the complex circle manifold, i.e.,
\begin{align*}
\mathrm{grad}_{\mathbf{\Theta}}\tilde f_{\Theta}
=
\mathrm{diag}\left(
\mathbf g_{\Theta}
-
\Re\left\{\mathbf g_{\Theta}\odot \boldsymbol{\vartheta}^*\right\}
\odot \boldsymbol{\vartheta}
\right),
\tag{27} \label{27}
\end{align*}
where $\odot$ denotes the Hadamard product.

The conjugate search direction and step size are updated using the same RCG and Armijo procedures as in the rotation-matrix subproblem. 
Given the search direction \(\boldsymbol{\Xi}_{\Theta}^{(t)}\) and step size \(\alpha^{(t)}\), the retraction onto \(\mathcal M_{\Theta}\) is performed by normalizing each diagonal entry, i.e.,
\begin{align*}
\mathbf{\Theta}^{(t+1)}
=
\mathrm{diag}\left(
\exp\left(
j\angle\left(
\mathrm{diag}\left(
\mathbf{\Theta}^{(t)}
+
\alpha^{(t)}
\boldsymbol{\Xi}_{\Theta}^{(t)}
\right)
\right)
\right)
\right).
\tag{28} \label{28}
\end{align*}

To ensure constraint satisfaction, the penalty parameters $\eta$ and $\lambda_j$ for $j \in \{2,3,4,5\}$ are increased upon violation within each subproblem. This ensures constraint satisfaction before proceeding to the next variable. The overall AO algorithm is summarized in Algorithm \ref{alg:AO_RCG}.

\begin{algorithm}[tb]
  \caption{Proposed AO Algorithm for Problem $\mathbf{P}1$}
  \label{alg:AO_RCG}
  \begin{algorithmic}[1]
    \Require Initial feasible points $\mathbf{R}^{(0)}$, $\mathbf{V}^{(0)}$, $\mathbf{\Theta}^{(0)}$, $\mathbf{u}_0^{(0)}$, and $\mathbf{w}^{(0)}$; penalty parameters $\eta^{\rm ini}$ and $\{\lambda_j^{\rm ini}\}_{j=2}^{5}$; scaling factor $\tau>1$; set outer iteration index $i=0$.
    \Ensure Optimized designs $\mathbf{R}^{\star}$, $\mathbf{V}^{\star}$, $\mathbf{\Theta}^{\star}$, $\mathbf{u}_0^{\star}$, and $\mathbf{w}^{\star}$.

    \Repeat
      \State Given $\mathbf{w}^{(i)}$, $\mathbf{V}^{(i)}$, $\mathbf{u}_0^{(i)}$, and $\mathbf{\Theta}^{(i)}$, set inner iteration index $t=0$ and initialize $\lambda_j\leftarrow\lambda_j^{\rm ini}$, $j=2,3,4,5$.
      \While{\eqref{10a}--\eqref{10c} or \eqref{10e} are not satisfied}
        \State Update $\mathbf{R}_{t+1}^{(i+1)}$ by solving $\mathbf{P}3.1$ via RCG method.
        \State If \eqref{10a}, \eqref{10b}, \eqref{10c}, or \eqref{10e} is violated, set the corresponding penalty parameter as
$\lambda_2\leftarrow\tau\lambda_2$, $\lambda_3\leftarrow\tau\lambda_3$, $\lambda_4\leftarrow\tau\lambda_4$, or $\lambda_5\leftarrow\tau\lambda_5$, respectively. Set $t\leftarrow t+1$.
      \EndWhile

      \State Given $\mathbf{w}^{(i)}$, $\mathbf{R}^{(i+1)}$, $\mathbf{u}_0^{(i)}$, and $\mathbf{\Theta}^{(i)}$, set inner iteration index $t=0$ and initialize $\lambda_j\leftarrow\lambda_j^{\rm ini}$, $j=2,3,4$.
       \While{\eqref{10a}--\eqref{10c} are not satisfied}
        \State Update $\mathbf{V}_{t+1}^{(i+1)}$ by solving $\mathbf{P}4$ via RCG method.
        \State If \eqref{10a}, \eqref{10b}, or \eqref{10c} is violated, set the corresponding penalty parameter as
        $\lambda_2\leftarrow\tau\lambda_2$, $\lambda_3\leftarrow\tau\lambda_3$, or $\lambda_4\leftarrow\tau\lambda_4$, respectively. Set $t\leftarrow t+1$.
      \EndWhile

      \State Given $\mathbf{w}^{(i)}$, $\mathbf{R}^{(i+1)}$, $\mathbf{V}^{(i+1)}$, and $\mathbf{\Theta}^{(i)}$, set inner iteration index $t=0$ and initialize $\lambda_j\leftarrow\lambda_j^{\rm ini}$, $j=2,3,4$.
      \While{\eqref{10a}--\eqref{10c} are not satisfied}
        \State Update $\mathbf{u}_{0,t+1}^{(i+1)}$ by solving $\mathbf{P}5$ via RCG method.
        \State If \eqref{10a}, \eqref{10b}, or \eqref{10c} is violated, set the corresponding penalty parameter as
        $\lambda_2\leftarrow\tau\lambda_2$, $\lambda_3\leftarrow\tau\lambda_3$, or $\lambda_4\leftarrow\tau\lambda_4$, respectively. Set $t\leftarrow t+1$.
      \EndWhile

      \State Given $\mathbf{w}^{(i)}$, $\mathbf{R}^{(i+1)}$, $\mathbf{V}^{(i+1)}$, and $\mathbf{u}_0^{(i+1)}$, set inner iteration index $t=0$ and initialize $\lambda_j\leftarrow\lambda_j^{\rm ini}$, $j=2,3,4$.
      \While{\eqref{10a}--\eqref{10c} are not satisfied}
        \State Update $\mathbf{\Theta}_{t+1}^{(i+1)}$ by solving $\mathbf{P}6$ via RCG method.
        \State If \eqref{10a}, \eqref{10b}, or \eqref{10c} is violated, set the corresponding penalty parameter as
        $\lambda_2\leftarrow\tau\lambda_2$, $\lambda_3\leftarrow\tau\lambda_3$, or $\lambda_4\leftarrow\tau\lambda_4$, respectively. Set $t\leftarrow t+1$.
      \EndWhile

      \State Given $\mathbf{R}^{(i+1)}$, $\mathbf{V}^{(i+1)}$, $\mathbf{u}_0^{(i+1)}$, and $\mathbf{\Theta}^{(i+1)}$, set inner iteration index $t=0$ and initialize $\eta\leftarrow\eta^{\rm ini}$.
      \While{The rank-one constraint \eqref{11d} is not satisfied}
        \State Update $\mathbf{D}_{t+1}^{(i+1)}$ by solving $\mathbf{P}2.1$.
        \State If \eqref{11d} is violated, set the penalty parameter as $\eta\leftarrow\tau\eta$. Set $t\leftarrow t+1$.
      \EndWhile

      \State Recover the digital beamformer $\mathbf{w}^{(i+1)}$ from $\mathbf{D}^{(i+1)}$.
      \State Update iteration index $i\leftarrow i+1$.
    \Until{The objective value of $\mathbf{P}1$ converges.}
  \end{algorithmic}
\end{algorithm}
\subsection{Convergence and Complexity Analysis}
\subsubsection{Convergence Analysis}
Let \(P^{(i)}=\|\mathbf w^{(i)}\|^2\) denote the transmit power after the \(i\)-th outer AO iteration. 
In each AO iteration, the rotation matrices \(\mathbf R\), the transmit polarization vectors \(\mathbf V\), the receive polarization vector \(\mathbf u_0\), and the DP-RIS phase shifts \(\boldsymbol{\Theta}\) are first updated with the digital beamformer fixed. 
Therefore, these updates do not directly change the transmit power. 
Instead, they are designed to improve the normalized constraint margins associated with the primary rate, secondary rate, and interference-temperature constraints, while satisfying the corresponding hardware constraints.
Specifically, at the $(i-1)$-th AO iteration, suppose that 
$(\mathbf R^{(i-1)},\mathbf V^{(i-1)},\mathbf u_0^{(i-1)},\boldsymbol{\Theta}^{(i-1)},\mathbf w^{(i-1)})$ is feasible for problem \(\mathbf{P}1\). 
During the updates of \(\mathbf R\), \(\mathbf V\), \(\mathbf u_0\), and \(\boldsymbol{\Theta}\), the beamforming \(\mathbf w^{(i-1)}\) is fixed, so the transmit power remains unchanged. 
Since these subproblems are solved while maintaining feasibility, \(\mathbf w^{(i-1)}\) is still feasible for the following beamforming update.
Given the updated variables 
\((\mathbf R^{(i)},\mathbf V^{(i)},\mathbf u_0^{(i)},\boldsymbol{\Theta}^{(i)})\), 
the digital beamforming subproblem is solved to minimize the transmit power. 
Since \(\mathbf w^{(i-1)}\) is a feasible solution to this subproblem, the optimized beamformer \(\mathbf w^{(i)}\) satisfies
\begin{align*}
P^{(i)}
=
\left\|\mathbf w^{(i)}\right\|^2
\le
\left\|\mathbf w^{(i-1)}\right\|^2
=
P^{(i-1)} .\tag{29} \label{29}
\end{align*}
Therefore, the transmit power sequence \(\{P^{(i)}\}\) is non-increasing over the AO iterations. Therefore, the transmit power sequence \(\{P^{(i)}\}\) is guaranteed to converge. Moreover, due to the non-convexity of problem \(\mathbf{P}1\), the obtained solution is generally a locally optimal solution.
\subsubsection{Computational Complexity}

We analyze the computational complexity of the proposed AO algorithm in terms of its main optimization blocks. 
Let \(I_{\rm AO}\) denote the number of outer AO iterations. 
Let \(I_{\rm D}\) denote the number of DC iterations for the digital beamforming subproblem, and let \(I_{\rm R}\), \(I_{\rm V}\), \(I_{\rm u}\), and \(I_{\Theta}\) denote the numbers of RCG iterations for updating \(\mathbf R\), \(\mathbf V\), \(\mathbf u_0\), and \(\boldsymbol{\Theta}\), respectively. 
For notational simplicity, we use \(L\) to denote the number of multipath components for each link.
For the digital beamforming update, the main complexity comes from solving the SDP/DC subproblem with an \(M\times M\) PSD matrix \(\mathbf D\). 
Following the standard complexity of interior-point methods, the complexity is
   $ \mathcal S_{\rm D}
    =
    \mathcal O\left(I_{\rm D}M^{4.5}\log(1/\epsilon)\right)$,
where \(I_{\rm D}\) is the number of DC iterations and \(\epsilon\) is the solution accuracy.
For the update of the rotation matrices, the main computational complexity arises from regenerating the channels affected by R and evaluating the smooth objective function and its Euclidean gradients. Specifically, regenerating the channels has a complexity of $\mathcal O(MNL+MKL)$, and computing the gradients requires $\mathcal O(MNK)$. Therefore, the overall complexity for updating \(\mathbf R\) is ${\mathcal S}_R= \mathcal O(I_R(MNL+MKL+MNK))$. For the RCG updates of $\mathbf V$, $\mathbf u_0$, and $\boldsymbol{\Theta}$, the relevant channel parameters can be precomputed and stored, and thus the complexity is mainly dominated by the gradient evaluation. Accordingly, the complexities for updating $\mathbf V$, $\mathbf u_0$, and $\boldsymbol{\Theta}$ are respectively given by $\mathcal S_{\rm V}=\mathcal O(I_{\rm V}MK)$, $\mathcal S_{\rm u}=\mathcal O(I_{\rm u})$ and $\mathcal S_{\Theta}=\mathcal O(I_{\Theta}NK)$, respectively. Therefore, the overall complexity of the proposed AO algorithm is given by $I_{\rm AO}(\mathcal S_{\rm D}+\mathcal S_{\rm V}+\mathcal S_{\rm u}+\mathcal S_{\rm \Theta})$.

\section{Low-Complexity Designs for Practical RA Implementation}
Although the proposed joint spatial-polarization design can fully exploit the rotation flexibility of RAs, it requires full continuous rotation control and optimization, which may lead to high hardware and computational complexity for large-scale arrays. 
For example, for an UPA equipped with $M$ antennas, RA control requires three independently controllable rotation DoFs for each antenna, and thus the number of rotation actuators, such as motors, may scale up to $3M$. Moreover, most existing RA optimization methods directly optimize the continuous rotation variables using iterative algorithms, such as RCG, SCA, or Frank-Wolfe, which requires high computational complexity. 
To reduce the hardware and computational complexity for practical deployment while preserving most spatial-polarization gains, we further develop two low-complexity RA designs for the rotation matrix optimization at the BS. Specifically, the first one reduces the number of independent rotation matrices by enforcing a subarray-wise shared rotation structure, while the second one replaces continuous rotation optimization with a discrete RA rotation codebook.
\subsection{Subarray-wise Shared Rotation Design} 
In the original full continuous rotation design, each antenna is assigned an independent rotation matrix \(\mathbf R_m\in \mathrm{SO}(3)\), which provides the highest orientation flexibility but also results in \(M\) independent rotation variables. 
To reduce the number of rotation variables, we consider a subarray-wise shared rotation design. 
Specifically, the BS antenna array is divided into $G$ disjoint subarrays, denoted by $\{\mathcal G_g\}_{g=1}^{G}$, satisfying
\begin{align*}
\bigcup_{g=1}^{G}\mathcal G_g = \{1,\ldots,M\},\tag{30} \label{30}
\end{align*}
where $\mathcal G_i\cap \mathcal G_j = \emptyset$ for any $ i\neq j $.
Moreover, antennas within the same subarray are constrained to share a common rotation matrix, i.e.,
\begin{align}
\mathbf R_m = \widetilde{\mathbf R}_g,\forall
m\in\mathcal G_g, \forall g=1,\ldots,G,
\tag{31} \label{31}
\end{align}
where $\widetilde{\mathbf R}_g\in\mathrm{SO}(3)$ denotes the shared rotation matrix of the $g$-th subarray.
Under this structure, the rotation search space is reduced from \(\mathrm{SO}(3)^M\) to \(\mathrm{SO}(3)^G\), where \(G\ll M\). 
From the hardware perspective, if each full 3D RA requires three independently controllable rotation DoFs, the number of rotation actuators can be reduced from up to \(3M\) to \(3G\). 
Therefore, the subarray-wise shared rotation design provides a practical tradeoff between performance and complexity.

The corresponding rotation optimization can be obtained by modifying the original RCG-based rotation update. 
Instead of optimizing all the rotation matrices \(\{\mathbf R_m\}_{m=1}^{M}\), we optimize the shared subarray rotation matrices \(\{\widetilde{\mathbf R}_g\}_{g=1}^{G}\). 
By substituting the shared-rotation constraint \(\mathbf R_m=\widetilde{\mathbf R}_g\), \(m\in\mathcal G_g\), into the smooth rotation objective in \(\mathbf{P}3.1\), the subarray-wise rotation optimization problem is formulated as
\begin{align}
\mathbf{P}7\;
\min_{\widetilde{\mathbf R}\in\mathcal M_R^{\rm sub}}
\;
\tilde f_R= \tilde f_1 + \lambda_2 \tilde f_2 + \lambda_3 \tilde f_3 + \lambda_4 \tilde f_4 + \lambda_5 \tilde f_5,
\tag{32}\label{32}
\end{align}
where \(\widetilde{\mathbf R}=\{\widetilde{\mathbf R}_g\}_{g=1}^{G}\) and 
\(\mathcal M_R^{\rm sub}=\mathrm{SO}(3)^G\). 
The main difference from \(\mathbf{P}3.1\) lies in the Euclidean gradient calculation. 
For each subarray \(g\), since the shared rotation matrix \(\widetilde{\mathbf R}_g\) affects all antennas in \(\mathcal G_g\), its Euclidean gradient is obtained by accumulating the gradients of the corresponding element-wise rotations, i.e.,
\begin{align}
\nabla_{\widetilde{\mathbf R}_g} \tilde f_R
=
\sum_{m\in\mathcal G_g}
\left.
\nabla_{\mathbf R_m} \tilde f_R
\right|_{\mathbf R_m=\widetilde{\mathbf R}_g},
\;\forall g=1,\ldots,G .
\tag{33}\label{33}
\end{align}
Then, problem \(\mathbf{P}7\) can be solved using the same RCG procedure as \(\mathbf{P}3.1\). 
The corresponding AO procedure is summarized in Algorithm~\ref{alg:shared_rotation}.
Compared with the full continuous RA design, this method preserves the continuous rotation optimization capability while significantly reducing the number of manifold variables.

\begin{algorithm}[tb]
  \caption{Subarray-wise Rotation Design Algorithm}
  \label{alg:shared_rotation}
  \begin{algorithmic}[1]
    \Require Initial feasible points $\widetilde{\mathbf R}^{(0)}$, $\mathbf{V}^{(0)}$, $\mathbf{\Theta}^{(0)}$, $\mathbf{u}_0^{(0)}$, and $\mathbf{w}^{(0)}$; subarray partition $\{\mathcal G_g\}_{g=1}^{G}$; penalty parameters $\eta^{\rm ini}$ and $\{\lambda_j^{\rm ini}\}_{j=2}^{5}$; scaling factor $\tau>1$; set iteration index $i=0$.
    \Ensure Optimized designs $\widetilde{\mathbf R}^{\star}$, $\mathbf{V}^{\star}$, $\mathbf{\Theta}^{\star}$, $\mathbf{u}_0^{\star}$, and $\mathbf{w}^{\star}$.

    \Repeat
      \State Given $\mathbf{w}^{(i)}$, $\mathbf{V}^{(i)}$, $\mathbf{u}_0^{(i)}$, and $\mathbf{\Theta}^{(i)}$, update $\widetilde{\mathbf R}^{(i+1)}$ by solving $\mathbf{P}7$ via RCG method.
      \State Set $\mathbf R_m^{(i+1)}=\widetilde{\mathbf R}_g^{(i+1)}$ for all $m\in\mathcal G_g$, $g=1,\ldots,G$.
      \State Update $\mathbf{V}^{(i+1)}$, $\mathbf{u}_0^{(i+1)}$, $\mathbf{\Theta}^{(i+1)}$, $\mathbf{D}^{(i+1)}$, and $\mathbf{w}^{(i+1)}$ following Algorithm~\ref{alg:AO_RCG}.
      \State Update iteration index $i\leftarrow i+1$.
    \Until{The objective value of $\mathbf{P}1$ converges.}
  \end{algorithmic}
\end{algorithm}
\subsection{Discrete RA Rotation Codebook}
Although the subarray-wise shared rotation design reduces the number of independent rotation variables, it still relies on continuous manifold optimization. 
To further reduce the algorithm complexity, we consider a discrete RA rotation codebook design. 
The key observation is that the directional gain of an RA mainly depends on the alignment between its boresight direction and the dominant propagation directions. 
Therefore, instead of searching over the continuous rotation space, we construct a small set of candidate boresight directions according to the system geometry.

Specifically, let \(\mathbf k_{\rm S}\) and \(\mathbf k_{\rm R}\) denote the unit direction vectors from the BS to the SR user and the DP-RIS, respectively. 
The candidate boresight codebook is constructed to include the SR-user direction, the DP-RIS direction, and multiple intermediate directions between them, i.e.,
\begin{align*}
\mathcal C_{\rm b}
=
\left\{
\frac{\alpha_q \mathbf k_{\rm S}+(1-\alpha_q)\mathbf k_{\rm R}}
{\left\|\alpha_q \mathbf k_{\rm S}+(1-\alpha_q)\mathbf k_{\rm R}\right\|}
\,\bigg|\,
\alpha_q\in\mathcal A
\right\},
\tag{34}\label{34}
\end{align*}
where \(\mathcal A\in[0,1]\) is a predefined set of weighting factors. 
By including \(\alpha_q=1\) and \(\alpha_q=0\), the codebook contains the SR-user direction and the DP-RIS direction, respectively, while the other values of \(\alpha_q\) generate intermediate directions between them.

For each candidate boresight direction $\mathbf c\in\mathcal C_{\rm b}$, we construct a deterministic rotation matrix $\mathbf R(\mathbf c)\in \mathrm{SO}(3)$. The first basis vector is set as
\begin{align*}
    \mathbf r_1=\mathbf c.\tag{35}\label{35}
\end{align*}
We then construct a second orthonormal direction by projecting the reference vertical polarization direction $\mathbf e_V=[0,0,1]^T$ onto the subspace orthogonal to $\mathbf r_1$, i.e.,
\begin{align*}
    \mathbf r_3
    =
    \frac{
    \left(\mathbf I_3-\mathbf r_1\mathbf r_1^T\right)\mathbf e_V
    }
    {
    \left\|
    \left(\mathbf I_3-\mathbf r_1\mathbf r_1^T\right)\mathbf e_V
    \right\|
    }. \tag{36}\label{36}
\end{align*}
Finally, the third basis vector is obtained as
\begin{align*}
    \mathbf r_2=\mathbf r_3\times \mathbf r_1,\tag{37}\label{37}
\end{align*}
which ensures that $\{\mathbf r_1,\mathbf r_2,\mathbf r_3\}$ forms a right-handed orthonormal basis.
Accordingly, the candidate rotation matrix is given by $\mathbf R(\mathbf c)=
    [\mathbf r_1,\mathbf r_2,\mathbf r_3]\in \mathrm{SO}(3)$. The resulting rotation codebook is defined as
\begin{align*}
\mathcal C_R \triangleq \{ \mathbf R(\mathbf c)\mid \mathbf c \in \mathcal C_b \}.\tag{38}\label{38}
\end{align*}

By constructing a finite rotation codebook, the continuous RCG-based rotation update can be replaced by a low-complexity discrete search. 
Specifically, within each AO iteration, the variables $\{\mathbf w,\mathbf V,\boldsymbol\Theta,\mathbf u_0\}$ are fixed, and the rotation matrices are updated one by one. The update of the $m$-th antenna rotation matrix is formulated as a discrete optimization problem over a predefined codebook $\mathcal C_R$, i.e.,
\begin{equation}
\mathbf{P}8 \quad\mathbf R_m^\star
=
\arg\min_{\mathbf R_m\in\mathcal C_R}
\tilde f_R(\mathbf R_1,\ldots,\mathbf R_M),
\tag{39}\label{39}
\end{equation}
where all other rotation matrices $\{\mathbf R_n\}_{n\neq m}$ are kept fixed.

In this way, the rotation update only requires \(M|\mathcal C_R|\) objective evaluations in each AO iteration, instead of performing iterative RCG-based optimization over the product manifold $\mathrm{SO}(3)^M$. 
Therefore, the discrete RA rotation codebook significantly reduces the rotation optimization complexity while preserving the desired directional gain of the antennas.

\begin{algorithm}[tb]
  \caption{Discrete RA Rotation Codebook Algorithm}
  \label{alg:codebook_rotation}
  \begin{algorithmic}[1]
    \Require Initial feasible points $\mathbf{R}^{(0)}$, $\mathbf{V}^{(0)}$, $\mathbf{\Theta}^{(0)}$, $\mathbf{u}_0^{(0)}$, and $\mathbf{w}^{(0)}$ with $\mathbf R_m^{(0)}\in\mathcal C_R$, $\forall m$; weighting set $\mathcal A$; penalty parameters $\eta^{\rm ini}$ and $\{\lambda_j^{\rm ini}\}_{j=2}^{5}$; scaling factor $\tau>1$; set iteration index $i=0$.
    \Ensure Optimized designs $\mathbf{R}^{\star}$, $\mathbf{V}^{\star}$, $\mathbf{\Theta}^{\star}$, $\mathbf{u}_0^{\star}$, and $\mathbf{w}^{\star}$.

    \State Construct the boresight codebook $\mathcal C_{\rm b}$ by \eqref{34}.
    \State Construct the rotation codebook $\mathcal C_R$ by \eqref{35}--\eqref{38}.

    \Repeat
      \State Given $\mathbf{w}^{(i)}$, $\mathbf{V}^{(i)}$, $\mathbf{u}_0^{(i)}$, and $\mathbf{\Theta}^{(i)}$.
      \For{$m=1,\ldots,M$}
        \State Update $\mathbf R_m^{(i+1)}$ by solving $\mathbf{P}8$ over the discrete codebook $\mathcal C_R$, with $\{\mathbf R_n\}_{n\neq m}$ fixed.
      \EndFor
      \State Update $\mathbf{V}^{(i+1)}$, $\mathbf{u}_0^{(i+1)}$, $\mathbf{\Theta}^{(i+1)}$, $\mathbf{D}^{(i+1)}$, and $\mathbf{w}^{(i+1)}$ following the corresponding steps in Algorithm~\ref{alg:AO_RCG}.
      \State Update iteration index $i\leftarrow i+1$.
    \Until{The objective value of $\mathbf{P}1$ converges.}
  \end{algorithmic}
\end{algorithm}

It is worth noting that, to preserve the monotonic convergence behavior of the AO framework, the initial rotation matrices should also be selected from the discrete codebook \(\mathcal C_R\). 
Otherwise, the first discrete rotation update may not guarantee a non-increasing value of the rotation objective \(\tilde f_R\) compared with an arbitrary continuous initialization. 

Moreover, the discrete RA rotation codebook is compatible with the subarray-wise shared rotation design. 
Specifically, when antennas in the same subarray share a common rotation matrix, the discrete search can be performed at the subarray level instead of the antenna level, i.e.,
\begin{align*}
\mathbf R_g^\star
=
\arg\min_{\mathbf R_g\in\mathcal C_R}
\tilde f_R
\left(
\{\mathbf R_m\}_{m=1}^{M}
\right),\tag{40}\label{40}
\end{align*}
where $\mathbf R_m = \mathbf R_g,\ \forall m\in\mathcal G_g$.
In this case, the number of objective evaluations in each rotation-update step is reduced from \(M|\mathcal C_R|\) to \(G|\mathcal C_R|\).

\section{Simulations} 
This section evaluates the performance of the proposed system through numerical simulations. The results validate the convergence of the proposed algorithm and demonstrate the superiority of the considered joint design. Unless otherwise specified, the default simulation parameters are set as follows. The BS and DP-RIS are located at $(0, 0, 10)$ m and $(100, 100, 10)$ m with $4\times4$ and $4\times8$ UPAs, respectively. The DP-RIS faces the $-y$-axis, and the antenna rotation is bounded by $\theta_{\max} = \pi/4$. All receivers are at a height of $1.5$ m. Relative to the $+x$-directed BS boresight $\mathbf{e}_B$, the SR and $K=2$ non-SR users are randomly distributed within $[150, 200]$ m and $[100, 200]$ m, at angles of $[25^\circ, 45^\circ]$ and $[-30^\circ, 30^\circ]$, respectively. Moreover, the receive polarization state of each non-SR user is fixed as vertical, given by $\mathbf{u}_k = [0, 1]^T$. We set $N_T=10$, $\sigma^2=-100$ dBm, ${I}_{\text{limit}}=-110$ dBm, $p=2$, and $\chi=0.1$. For all links, the numbers of paths are identical, i.e., $L_{BR}=L_{BU,i}=L_{RU,i}=4$. The LoS ($l=0$) path loss exponents are $\alpha_{0}^{BR}=\alpha_{0}^{RU,0}=2.0$, $\alpha_{0}^{BU,0}=3.5$, $\alpha_{0}^{BU,i}=3.7$, and $\alpha_{0}^{RU,i}=2.4$ ($i \ge 1$). For all NLoS paths ($l \ge 1$), the exponent is uniformly set to $\alpha_l^{(\cdot)}=4.0$~\cite{goldsmith2005wireless}.
Moreover, the SDP-based subproblems are solved using CVX, while the RCG-based manifold optimization subproblems are implemented using Manopt~\cite{boumal2014manopt}.

\subsection{Performance Comparison with Benchmark Schemes}
To verify the effectiveness of the proposed joint spatial-polarization design, we compare it with four benchmark schemes. For fair comparison, all schemes are evaluated under the same rate requirements in all schemes. The schemes differ in the use of antenna rotation, transmit polarization optimization, and DP-RIS phase optimization. Specifically, the benchmark schemes and their abbreviations are defined as follows:
\begin{itemize}
\item \textit{Baseline 1} (DBF w/ Random DP-RIS): Digital beamforming (DBF) with fixed polarization at the BS and random DP-RIS phase shifts. \looseness-1
\item \textit{Baseline 2} (Polarforming w/ Random DP-RIS): DBF with optimized polarization (polarforming) at the BS and random DP-RIS phase shifts.
\item \textit{Baseline 3} (DBF w/ Opt. DP-RIS): DBF with fixed polarization at the BS and optimized DP-RIS phase shifts. \looseness-1
\item \textit{Baseline 4} (Polarforming w/ Opt. DP-RIS): Polarforming at the BS and optimized DP-RIS phase shifts, but without antenna rotation.
\end{itemize}
\begin{figure}[t!]
  \centering
  \includegraphics[width=0.48\textwidth]{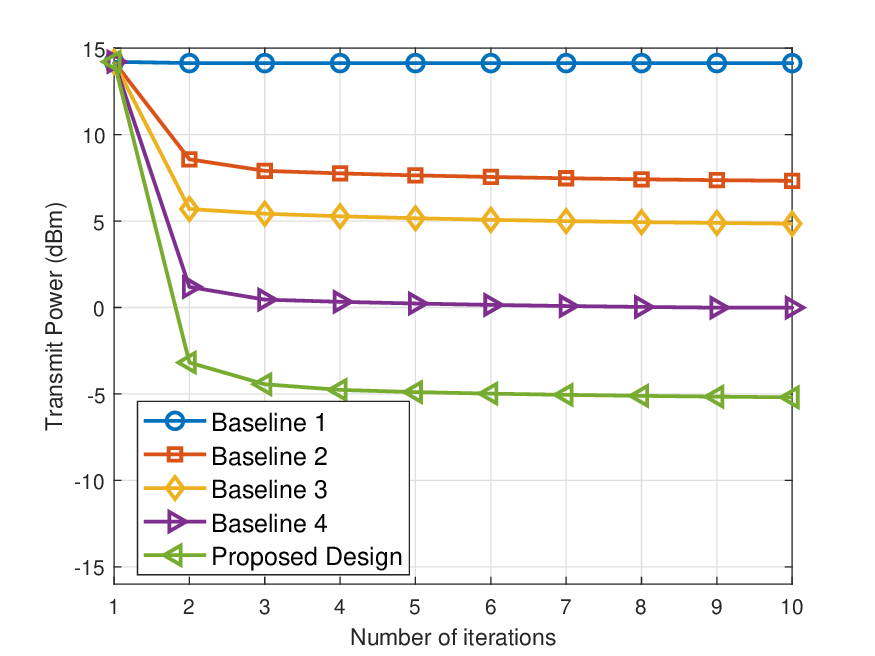}
  \caption{Convergence behavior of different schemes}
  \label{figure1}
\end{figure}

\begin{figure}[t!]
  \centering
  \includegraphics[width=0.48\textwidth]{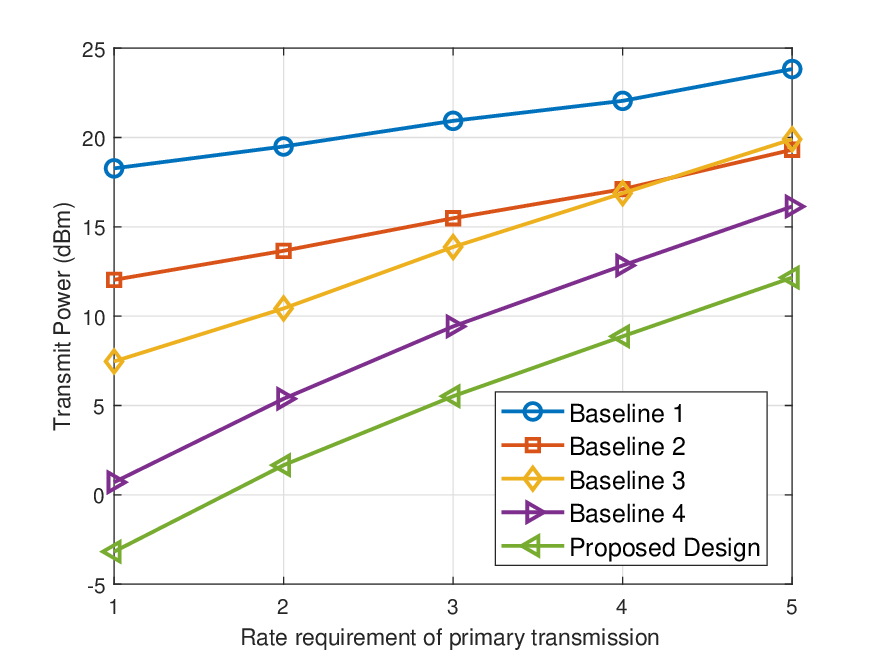}
  \caption{Transmit power versus $\bar{R}_s$, with $\bar{R}_c = 0.02$ bps/Hz}
  \label{figure2}
\end{figure}
Fig. \ref{figure1} illustrates the convergence behavior of different schemes, with $\bar{R}_s=1$ bps/Hz and $\bar{R}_c=0.02$ bps/Hz. Except for Baseline 1, which does not involve alternating optimization, all schemes monotonically reduce the required transmit power and converge within roughly $10$ iterations. This fast convergence demonstrates the effectiveness and robustness of the proposed alternating optimization framework.

\begin{figure}[t!]
  \centering
  \includegraphics[width=0.48\textwidth]{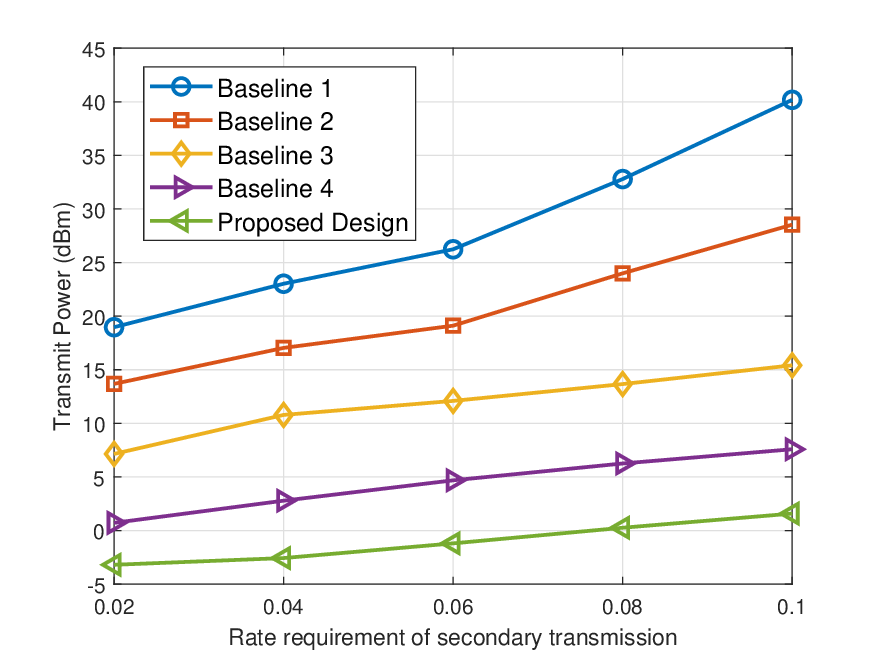}
  \caption{Transmit power versus $\bar{R}_c$, with $\bar{R}_s = 1$ bps/Hz}
  \label{figure3}
\end{figure}
\begin{figure}[t!]
  \centering
  \includegraphics[width=0.48\textwidth]{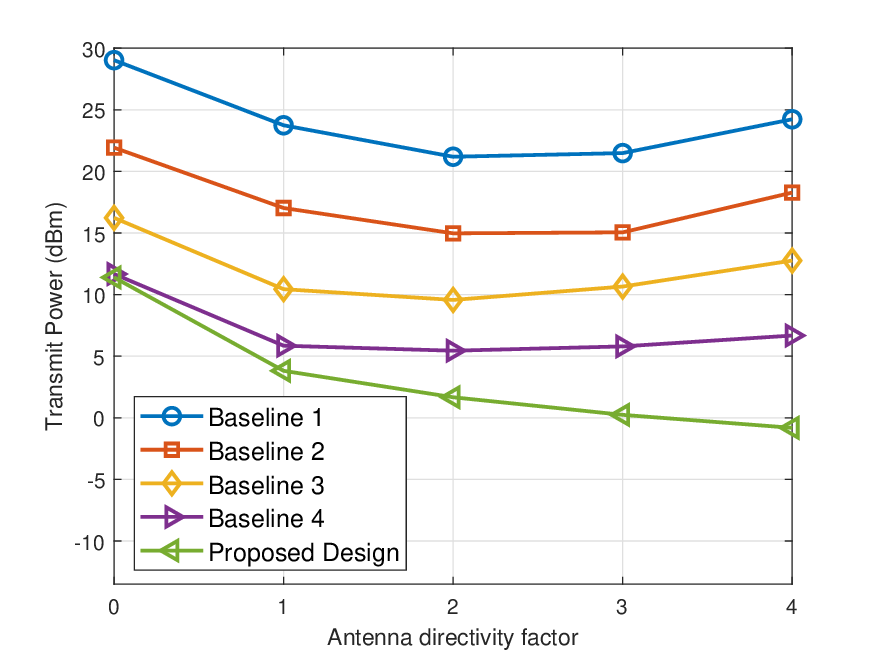}
  \caption{Transmit power versus $p$, with $\bar{R}_s = 2$ and $\bar{R}_c = 0.02$ bps/Hz}
  \label{figure4}
\end{figure}

Fig. \ref{figure2} shows the required transmit power versus the primary rate requirement $\bar{R}_s$, with $\bar{R}_c = 0.02$ bps/Hz. As $\bar{R}_s$ increases, all schemes require higher transmit power to satisfy the primary rate requirement. The proposed design achieves the lowest transmit power, and its reduction compared with Baseline 4 highlights the benefit of jointly optimizing antenna rotation and polarization states for polarization matching. Moreover, a crossover is observed between Baselines 2 and 3.
At low $\bar{R}_s$, Baseline 3 performs better because optimized DP-RIS phases are crucial for supporting the secondary transmission. As $\bar{R}_s$ grows, the increased transmit power satisfies the secondary rate requirement, while the primary rate constraint becomes dominant, making polarization matching at the BS more important and allowing Baseline 2 to surpass Baseline 3.\looseness-1

Fig. \ref{figure3} illustrates the required transmit power versus the secondary rate requirement $\bar{R}_c$, with $\bar{R}_s = 1$ bps/Hz. As expected, all schemes require higher transmit power as $\overline{R}_{c}$ increases to meet the more stringent secondary rate requirement. Notably, Baselines 1 and 2 require sharply increased transmit power as $\bar{R}_c$ grows. This is because random DP-RIS phases cannot provide effective passive beamforming gain, forcing the BS to increase its transmit power to compensate for the severe double fading of the cascaded backscatter link. In contrast, Baselines 3 and 4, which optimize the DP-RIS phase shifts, exhibit a more gradual increase in transmit power. This indicates that DP-RIS phase optimization provides effective passive beamforming gain and helps improve the cascaded backscatter link as the secondary rate requirement increases. Furthermore, Baseline 4 achieves lower transmit power than Baseline 3, showing that polarforming can mitigate polarization mismatch and further improve the secondary transmission.
Finally, the proposed joint design consistently achieves the minimum power consumption among all schemes. 

Fig. \ref{figure4} illustrates the required transmit power versus the antenna directivity factor $p$. For Baselines 1--4 with fixed antenna orientations, the transmit power initially decreases but eventually increases as $p$ grows. This trend reveals a physical limitation of fixed directional antennas. A larger $p$ provides a higher peak directional gain, but it also makes the radiation pattern more concentrated around the fixed boresight direction. Since the antennas cannot rotate, their boresight directions may not be well aligned with the desired direction. As the radiation pattern becomes more directional, this angular mismatch reduces the antenna gains toward the SR-user and DP-RIS directions. When $p$ becomes large, the loss caused by this misalignment offsets the peak gain improvement, leading to the increased transmit power observed in the fixed-orientation baselines. 
In contrast, as $p$ increases, the proposed design effectively exploits the high directivity gain through antenna rotation, leading to a monotonic decrease in the required transmit power and highlighting the importance of joint antenna rotation and polarization matching.
\subsection{Performance of Low-Complexity RA Designs}
We further evaluate the proposed low-complexity RA designs. The performance of the proposed low-complexity RA designs depends on whether the reduced rotation search space can provide a good approximation to the full continuous rotation design. 
This is mainly affected by the spatial distribution of the DP-RIS and users, as well as the antenna directivity factor $p$. 
When the SR user and the DP-RIS are located in similar angular directions with respect to the BS, the direct SR link and the RIS-assisted secondary link can benefit from similar antenna orientations. Therefore, a shared rotation design or a small codebook is sufficient to represent the dominant antenna orientation. In contrast, when the SR-user direction moves away from the DP-RIS direction, the RA orientation needs to balance the direct SR link, the RIS-assisted backscatter link, and the interference leakage toward non-SR users, making low-complexity rotation design more challenging. 
Moreover, a larger directivity factor $p$ leads to a narrower radiation pattern and hence increases the sensitivity to rotation mismatch.

Therefore, we evaluate the proposed low-complexity designs by considering different SR-user distributions and antenna directivity factors. Specifically, the SR user is located at a fixed distance of $200$ m from the BS, while its angular location with respect to the BS is randomly generated within one of the regions in $\{[30^\circ,45^\circ], [15^\circ,30^\circ], [0^\circ,15^\circ], [-15^\circ,0^\circ], [-30^\circ,-15^\circ]\}$.
The DP-RIS location and the receiver height are kept unchanged, so that the DP-RIS remains in the \(45^\circ\) direction with respect to the BS. 
This setup allows us to examine how the low-complexity designs perform when the SR-user direction gradually moves away from the DP-RIS direction. 
The rate requirements are set as \(\bar R_s=2\) bps/Hz for the primary transmission and \(\bar R_c=0.02\) bps/Hz for the secondary transmission.

\begin{figure}[t!]
  \centering
  \includegraphics[width=0.48\textwidth]{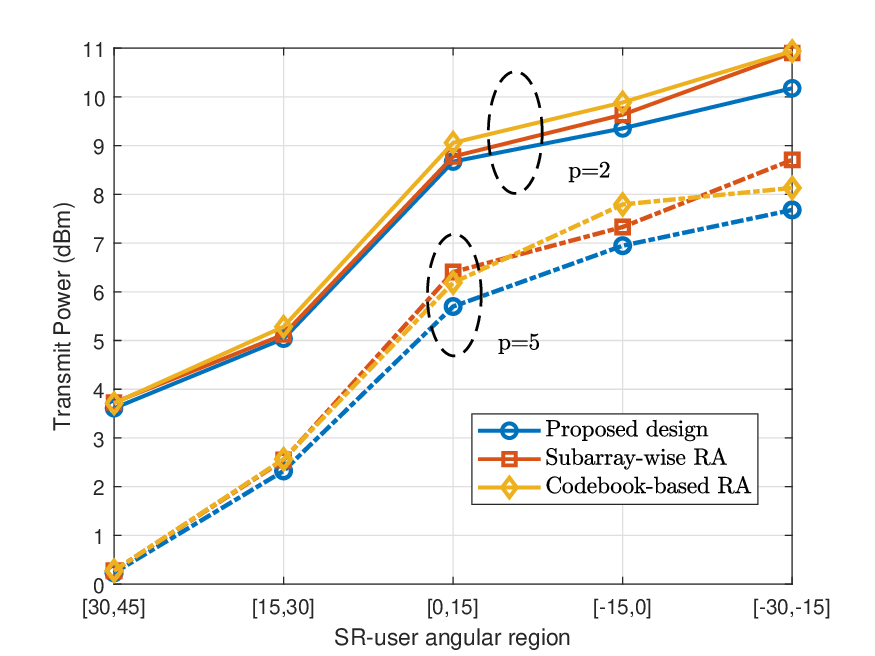}
  \caption{Transmit power versus the SR-user direction, with $\bar{R}_s = 2$ and $\bar{R}_c = 0.02$ bps/Hz}
  \label{figure8}
\end{figure}

Fig.~\ref{figure8} shows the required transmit power versus the SR-user angular region. 
The lower three curves correspond to \(p=5\), while the upper three curves correspond to \(p=2\). 
For the subarray-wise shared rotation design, the BS array is divided into two subarrays, and thus only two shared rotation matrices need to be optimized. 
For the discrete RA rotation design, we adopt a three-candidate codebook with \(\mathcal A=\{0,0.5,1\}\), corresponding to the SR-user direction, the DP-RIS direction, and their middle direction, respectively.

It can be observed that the required transmit power generally increases as the SR-user angular region moves from \([30^\circ,45^\circ]\) to \([-30^\circ,-15^\circ]\). This is mainly because the SR user gradually moves away from the DP-RIS direction. As a result, the RIS-assisted cascaded link experiences larger path loss, and more transmit power is needed to satisfy the rate requirement of the secondary transmission. In addition, the curves with $p=5$ require lower transmit power than those with $p=2$, showing that a larger directivity factor can provide additional directional gain when antenna rotation is available. Furthermore, when the SR-user direction is close to the DP-RIS direction, the two low-complexity RA designs achieve nearly the same performance as the full continuous RA design. This indicates that, in this case, a small number of shared rotations or codebook candidates can provide a good approximation to the full rotation design. As the SR-user direction moves farther away from the DP-RIS direction, the performance gaps gradually become larger. This is because the favorable antenna orientations become more diverse when the direct SR link and the RIS-assisted cascaded link correspond to more separated spatial directions. The full continuous RA design can adjust the rotation matrix of each antenna more flexibly, whereas the subarray-wise shared rotation design and the discrete rotation codebook design use a reduced rotation search space. Nevertheless, the performance loss of the proposed low-complexity designs remains limited. For example, when $p=2$, the maximum gap is about $0.7$ dBm. This shows that the proposed low-complexity RA designs can preserve most of the transmit-power reduction achieved by the full continuous RA design while reducing the rotation control and optimization complexity.

\begin{figure}[t!]
  \centering
  \includegraphics[width=0.48\textwidth]{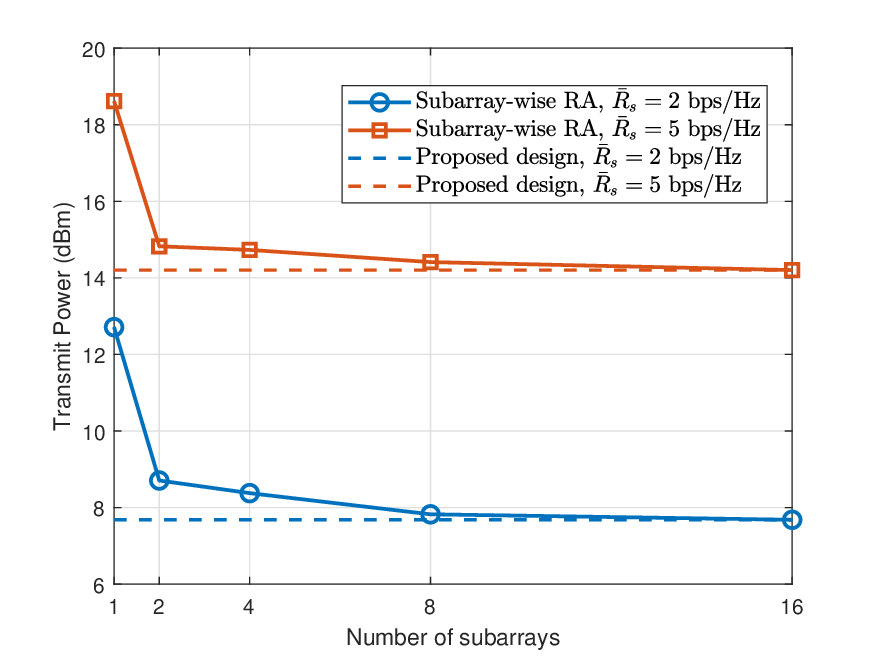}
  \caption{Transmit power versus the number of subarrays, with $p = 5$ and $\bar{R}_c = 0.02$ bps/Hz}
  \label{figure9}
\end{figure}

\begin{figure}[t!]
  \centering

  \subfloat[$\bar{R}_s = 2$ and $\bar{R}_c = 0.02$ bps/Hz]{
    \includegraphics[width=0.48\textwidth]{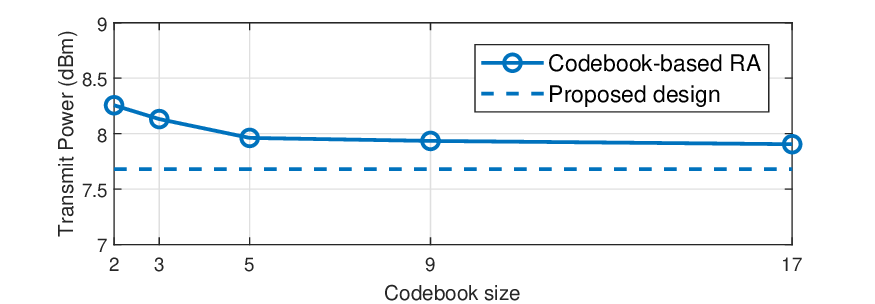}
  }

  \vspace{-0.3cm}

  \subfloat[$\bar{R}_s = 5$ and $\bar{R}_c = 0.02$ bps/Hz]{
    \includegraphics[width=0.48\textwidth]{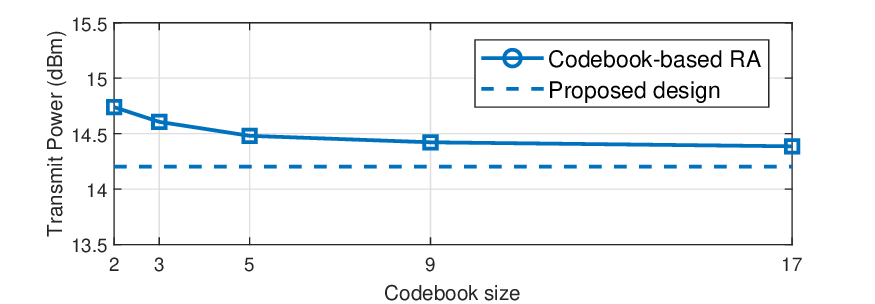}
  }

  \caption{Transmit power versus codebook size, with $p = 5$}
    \label{figure10}
\end{figure}


Fig.~\ref{figure9} further evaluates the impact of the number of subarrays on the subarray-wise shared rotation design. The total BS array is divided into $G$ subarrays, where $G\in\{1,2,4,8,16\}$. Increasing $G$ provides more independent rotation matrices and therefore improves the rotation flexibility, but it also increases the number of rotation variables and the associated complexity. It can be observed that increasing $G$ from $1$ to $2$ leads to a clear reduction in the required transmit power. This is because, when all antennas share a single rotation matrix, the array has limited flexibility to support the direct SR link and the RIS-assisted cascaded link at the same time. With two subarrays, the BS can use two shared rotations to better adapt to these two main directions, which brings most of the performance improvement. When $G$ increases beyond $2$, the transmit-power reduction gradually slows down and the performance approaches that of the full continuous RA design. This observation suggests that the number of shared rotations does not need to be very large. As long as the shared rotations can cover the main desired directions, the subarray-wise design can achieve comparable performance to the full continuous design. Therefore, the subarray-wise shared rotation design provides a practical balance between system performance and complexity.

Fig.~\ref{figure10} illustrates the required transmit power versus the codebook size for the discrete RA rotation design. Two rate requirements are considered to examine the effect of the codebook size under different rate requirements of the primary transmission. As the codebook size increases, the required transmit power decreases because more candidate rotations are available and the discrete design can better approximate the full continuous RA design. This shows that adding more candidate rotations improves the approximation capability of the discrete design. However, once the main desired spatial directions are represented in the codebook, further increasing the codebook size brings only limited additional improvement. Therefore, a small rotation codebook can preserve most of the spatial gain while reducing the rotation optimization complexity.\looseness-1
\section{Conclusion}
In this paper, we have investigated a DP-RIS-empowered SR system, where the BS antennas support both polarization reconfiguration and antenna rotation. To exploit these capabilities, we have formulated a transmit power minimization problem by jointly optimizing beamforming, antenna rotation, polarization states, and DP-RIS phase shifts. We have further solved the resulting non-convex problem using an AO algorithm with SDP, DC programming, and RCG methods. To reduce the hardware and computational complexity of RA implementation, we have further developed two low-complexity rotation designs, namely the subarray-wise shared rotation design and the discrete RA rotation codebook design. Simulation results have shown that the proposed joint spatial-polarization design effectively enhances the primary and secondary transmissions while mitigating interference to non-SR users. 
Moreover, the proposed low-complexity designs preserve most of the performance gain of the full continuous rotation design with significantly reduced rotation complexity.

\bibliographystyle{IEEEtran}
\bibliography{main.bib}
\end{document}